\documentclass[aps,reprint,nofootinbib,nobibnotes,notitlepage,superscriptaddress,twocolumn,prl,
 amsmath,amssymb
]{revtex4-2}

\usepackage[caption=false]{subfig}
\usepackage{braket}
\usepackage{float}
\usepackage{lipsum}
\usepackage{graphicx}
\usepackage{dcolumn}
\usepackage{bm}
\usepackage{natbib}
\usepackage{hyperref}
\hypersetup{
	colorlinks = true,
    linkcolor = Red,
    urlcolor  = Red,
    citecolor = Red
}

\usepackage{microtype}

\usepackage{verbatim}
\usepackage[amssymb]{SIunits}
\usepackage{tabularx}
\usepackage[dvipsnames]{xcolor}
\usepackage{wasysym}

\usepackage{tikz}

\def\be{\begin{equation}}
\def\ee{\end{equation}}

\usepackage{color}
\definecolor{darkgreen}{RGB}{0,120,0}

\definecolor{darkgreen}{RGB}{0,120,0}

\newcommand{\resub}[1]{{#1}}

\newcommand{\hMpc}{h\,\mathrm{Mpc}^{-1}}

\newcommand{\delD}[1]{(2\pi)^3\delta_\mathrm{D}\left({#1}\right)}

\newcommand{\av}[1]{\left\langle{#1}\right\rangle} 

\newcommand{\vk}{\vec k}
\newcommand{\hk}{\hat{\vec k}}

\newcommand{\vs}{\vec s}

\newcommand{\vx}{\vec x}
\newcommand{\hx}{\hat{\vec x}}

\newcommand{\F}{\mathcal{F}}

\newcommand{\hs}{\hat{\vec s}}
\newcommand{\tjo}[3]{\begin{pmatrix} {#1} & {#2} & {#3}\\ 0 & 0 & 0\end{pmatrix}}
\newcommand{\tj}[6]{\begin{pmatrix} {#1} & {#2} & {#3}\\ {#4} & {#5} & {#6}\end{pmatrix}}

\renewcommand{\L}{\Lambda}
\renewcommand{\P}{\mathcal{P}}

\def\beq{\begin{eqnarray}}
\def\eeq{\end{eqnarray}}
\def\k{\textbf{k}}
\def\bfk{\textbf{k}}

\let\vec\mathbf

\usepackage{empheq}

\def\aap{\rm{A\&A}}

\begin{document}

\title{Do the CMB Temperature Fluctuations Conserve Parity?}

\author{Oliver~H.\,E.~Philcox}
\email{ohep2@cantab.ac.uk}
\affiliation{Center for Theoretical Physics, Department of Physics,
Columbia University, New York, NY 10027, USA}
\affiliation{Simons Society of Fellows, Simons Foundation, New York, NY 10010, USA}

\begin{abstract}  
    Observations of the Cosmic Microwave Background (CMB) have cemented the notion that the large-scale Universe is both statistically homogeneous and isotropic. But is it invariant also under reflections? To probe this we require parity-sensitive statistics: for scalar observables, the simplest is the \resub{trispectrum}. We make the first measurements of the parity-odd \resub{scalar CMB}, focusing on the large-scale ($2<\ell<510$) temperature anisotropies measured by \textit{Planck}. This is facilitated by new \resub{quasi}-maximum-likelihood estimators for binned correlators, which account for mask convolution and leakage between even- and odd-parity components, and achieve \resub{ideal} variances within $\approx 20\%$. We perform a blind test for parity violation by comparing a $\chi^2$ statistic from \textit{Planck} to theoretical expectations, using two suites of simulations to account for the possible likelihood non-Gaussianity and residual foregrounds. We find consistency at the \resub{$\approx 0.4\sigma$} level, \resub{yielding no evidence for novel early-Universe phenomena}. The measured trispectra \resub{allow for a wealth of new physics to be constrained; here, we use them to constrain eight primordial models, including Ghost Inflation, Cosmological Collider scenarios, and Chern-Simons gauge fields. We find no signatures of new physics}, with a maximal detection significance of $2.0\sigma$. 
    \resub{Our results also indicate that the recent parity excesses seen in the BOSS galaxy survey are not primordial in origin, given that the CMB dataset contains roughly $250\times$ more primordial modes, and is far easier to interpret, given the} linear physics, Gaussian statistics, and accurate mocks. Tighter CMB constraints can be wrought by including smaller scales (though rotational invariance washes out the flat-sky limit) and adding polarization data. 
\end{abstract}

\maketitle

\noindent Since their inception in the mid-1960s, analyses of the Cosmic Microwave Background (CMB) have revolutionized modern physics, most notably through the associated development of a cosmological model \citep[e.g.,][]{WMAP:2003elm,2020A&A...641A...6P}. Though early work focused on the CMB frequency spectrum \citep[e.g.,][]{Fixsen:1996nj}, most contemporary experiments consider the spatial fluctuations, both in temperature and polarization \citep[e.g.,][]{ACT:2020gnv,BICEP2:2015xme,SPT-3G:2014dbx}. Through the measurement of polyspectra, including the power spectrum, bispectrum, and trispectrum, these can be used to constrain the physics of the early and late Universe, placing constraints on a wide variety of models. Upcoming experiments will significantly tighten these bounds, with much of the new information coming from polarization anisotropies at high multipoles \citep{SimonsObservatory:2018koc,CMB-S4:2016ple}. Given this push to smaller scales, it is interesting to ask whether this is sufficient for all models of interest; equivalently, have we exhausted all the information in the large-scale temperature fluctuations probed by full-sky experiments such as \textit{Planck}?

Cosmological parity symmetry provides an intriguing counterexample. In particle physics, it has long been known that weak interactions do not preserve parity (hereafter $\mathbb{P}$) symmetry \citep[e.g.,][]{PhysRev.104.254}; cosmologically, most interactions of interest are gravitational, and thus $\mathbb{P}$-invariant. In the early Universe, this is not the case, and there are theoretical hints that parity violation could accompany phenomena such as baryogenesis (via some leptogenesis mechanism) \citep{Alexander:2011hz,Alexander:2007gj,Alexander:2016hxk,Sakharov:1967dj}. If this occurs during inflation (with examples involving massive exchange particles \citep{Arkani-Hamed:2015bza} and modified gravity \citep{Alexander:2011hz}), signatures would naturally be left in the distribution of matter and gravitational waves. In the latter case, $\mathbb{P}$-violating power spectra such as $TB$ and $EB$ would be generated,\footnote{These can also be generated by anisotropic (parity-conserving) inflationary models, though this generates only off-diagonal contributions \citep[e.g.,][]{Watanabe:2009ct,Watanabe:2010bu,Nicolis:2020rqz,Gumrukcuoglu:2010yc,Bartolo:2014hwa}. \resub{Gravitational wave effects can also generate scalar four-point functions via ``fossil'' effects \citep[cf.,][]{Masui:2017fzw}, though these are constrained from their CMB $B$-mode signatures.}} or higher-order bispectra such as $TTB$ \citep[e.g.,][]{Lue:1998mq,Gluscevic:2010vv,Kamionkowski:2010rb,Shiraishi:2013kxa,Shiraishi:2013vha,Shiraishi:2012sn,Shiraishi:2011st,Bartolo:2018elp,Bartolo:2014hwa,Molinari:2018wvn}. For scalar observables, such as temperature fluctuations, $\mathbb{P}$-violating physics can only be seen in a temperature trispectrum or above, due to the equivalence of parity-transformations and three-dimensional rotations for low order statistics (first discussed for galaxy survey contexts in \citep{Cahn:2021ltp}). This has been noted in previous works \citep[e.g.,][]{Shiraishi:2016mok,Philcox:2022hkh,Cabass:2022oap,Alexander:2006mt,Coulton:2023oug}, but never explicitly searched for in CMB data.\footnote{Note that new physics models generically modify the scalar and tensor CMB in different ways, thus it remains useful to look at the temperature four-point function after a non-detection of the $B$-mode two-point functions.}

Determining whether the CMB obeys parity symmetry is a topic of great current relevance. Recently, tentative evidence for $\mathbb{P}$-violation has been reported using CMB birefringence \citep{Diego-Palazuelos:2022dsq,Eskilt:2022wav,Eskilt:2022cff}, as well as the large-scale distribution of galaxies \citep{Philcox:2022hkh,Hou:2022wfj}, with the latter using the approach first proposed in \citep{Cahn:2021ltp}. Given that systematic effects such as unaccounted-for cosmic dust \citep[e.g.,][]{Clark:2021kze} or unjustified analysis choices (e.g., a Gaussian likelihood or mocks of insufficient fidelity) could be an alternative explanation for such results, full confirmation requires an independent probe. In this \textit{Letter}, \resub{we perform the first tests of scalar $\mathbb{P}$-violation in the CMB by measuring} the four-point correlator of the CMB temperature anisotropies, using \resub{methods similar} to those of other primordial non-Gaussianity analyses (see \citep{Planck:2019kim} and references therein), as well as new tools for estimating the correlators themselves \citep{Philcox:2023uwe}. Since this is a scalar observable, it cannot be used to directly constrain birefringence via axion-photon interactions \citep[e.g.,][]{Carroll:1998zi,Turner:1987bw,Gluscevic:2010vv,Alexander:2016hxk}); these occur at late-times, and affect only photon polarization. However, it is a direct probe of the primordial density perturbations, and is thus sensitive to potential early Universe processes that could source parity violation in the distribution of galaxies. Indeed, the CMB is likely a stronger probe, given its large signal-to-noise and almost-Gaussian statistics.

Given this novel dataset, we may ask two questions: (1) do we see \textit{any} evidence of parity violation in the large-scale \textit{Planck} temperature map? (2) what bounds can we place on physical models of parity violation? The latter can be compared to constraints from galaxy surveys \citep{Cabass:2022oap}, with the current results benefiting from a much larger maximum scale ($\ell\approx3$), the cosmic-variance limit of the CMB, and the well understood statistics and covariances. Below, we present this analysis, with additional details given in the Supplementary Material. To avoid confirmation bias, the \resub{entire} (\href{https://GitHub.com/OliverPhilcox/PolyBin/planck_public}{public}) analysis pipeline (\resub{including trispectrum estimation, model computation, and null test pipeline}) was developed and tested \resub{on realistic simulations} before the \textit{Planck} data were analyzed.

\section*{Theoretical Framework}
\noindent The reduced CMB trispectrum, $t^{\ell_1\ell_2}_{\ell_3\ell_4}(L)$, is defined as \beq\label{eq: reduced-Tl-def}
    \bigg\langle{\prod_{i=1}^4a_{\ell_im_i}}\bigg\rangle_c&\equiv&\sum_{LM}(-1)^Mw^{L(-M)}_{\ell_1\ell_2m_1m_2}w^{LM}_{\ell_3\ell_4m_3m_4}t^{\ell_1\ell_2}_{\ell_3\ell_4}(L)\nonumber\\
    &&\,+\,\text{23 perms.}
\eeq
\citep{Regan:2010cn,Philcox:2023uwe}, where $a_{\ell m}$ is the primary CMB temperature fluctuation, defining the weights 
\beq\label{eq: weight-def}
    w^{LM}_{\ell_1\ell_2m_1m_2} &\equiv& \sqrt{\frac{(2\ell_1+1)(2\ell_2+1)(2L+1)}{4\pi}}\\\nonumber
    &&\,\times\,\tj{\ell_1}{\ell_2}{L}{m_1}{m_2}{-M}\tj{\ell_1}{\ell_2}{L}{-1}{-1}{2}
\eeq
(choosing spins as in \citep{Philcox:2023uwe} to avoid parity-odd cancellations). Here, the $3\times 2$ matrices are Wigner $3j$ symbols, and the trispectrum is indexed by the multipoles of four sides, $\ell_i$, and a diagonal, $L$, with $\{\ell_1,\ell_2,L\}$ and $\{\ell_3,\ell_4,L\}$ obeying triangle conditions. Trispectra with even (odd) $\ell_1+\ell_2+\ell_3+\ell_4$ probe the parity-even (parity-odd) contributions to the CMB perturbations; we consider the latter in this work.

The CMB temperature fluctuations are a direct probe of the underlying primordial curvature perturbation, $\zeta(\vk)$:
\beq
    a_{\ell m} \equiv 4\pi i^\ell \int_{\vk}\zeta(\vk)\mathcal{T}_\ell(k)Y_{\ell m}^*(\hk),
\eeq
where $Y_{\ell m}$ is a spherical harmonic, $\int_{\vk}\equiv \int d^3k/(2\pi)^3$, and $\mathcal{T}_\ell$ is the usual transfer function, encoding radiation physics. It follows that the CMB trispectrum is linearly related to the curvature trispectrum, $T_\zeta$: as such, it can be used to place constraints on various models of inflationary parity violation. 


\section*{Data Analysis}
\noindent Our dataset is the full-mission \textit{Planck} 2018 SMICA component-separated temperature map \citep{Planck:2019kim,Planck:2018yye}.\footnote{Available at \href{http://pla.esac.esa.int}{pla.esac.esa.int}.} We additionally use a set of 300 FFP10 simulations \citep{Planck:2015mis,Planck:2018lkk} to validate our pipeline; these are generated at a cosmology similar to \citep{Planck:2015fie} and processed with the same SMICA pipeline. To model the trispectrum noise properties, we generate a suite of 1000 Gaussian random field (hereafter GRF) maps at a \resub{\textsc{healpix} $N_{\rm side}=2048$ \citep{Gorski:2004by}} with the \textit{Planck} 2018 best-fit cosmology and noise properties \citep{Planck:2019kim}. \resub{To begin our analysis, we downgrade all maps to $N_{\rm side}=256$ (for computational efficiency),} then modulated them by a linear filter, denoted $\mathsf{S}^{-1}$, similar to the approach of \citep[][\S9.1]{Planck:2015zfm} and \citep{2015arXiv150200635S}. This \resub{applies the common sky mask} described in \citep{Planck:2019kim} (with $10$-arcminute smoothing, yielding $f_{\rm sky}=0.79$) to null the brightest foreground regions and linearly inpaints small holes (those containing fewer than $40$ pixels), as described in \citep{Gruetjen:2015sta}. 
Maps are then filtered in harmonic-space \resub{(at $N_{\rm side}=256$)} using a $C_\ell^{TT}+N_\ell$ transfer function, where $C_\ell^{TT}$ is the theoretical power spectrum evaluated at the best-fit cosmology of \citep{2020A&A...641A...6P}, and the noise spectrum $N_\ell$ is extracted from the SMICA half-mission maps.

Given the data, the $\mathsf{S}^{-1}$ weighting scheme, and the \textit{Planck} beam (including the pixel window function), correlators are obtained using the \textsc{PolyBin} code (see \citep{Philcox:2023uwe} for discussion and verification).\footnote{Available at \href{https://github.com/oliverphilcox/PolyBin}{GitHub.com/OliverPhilcox/PolyBin} \citep{PolyBin}.} This implements the binned trispectrum estimator, $\widehat{t}$, from a quartic estimator ($\widehat{t}^{\rm num}$, with residual disconnected contributions subtracted using $100$ GRF simulations) and a data-independent normalization matrix $\mathcal{F}$:
\beq\label{eq: polybin-eq}
    \widehat{t}(\vec b,B) = \sum_{\vec b',B'}\mathcal{F}^{-1}(\vec b,B;\vec b',B')\widehat{t}^{\rm num}(\vec b',B'),
\eeq
where $\vec b,B$ are bins in $\{\ell_i\}$ and $L$ respectively (cf.\,\citep{2021PhRvD.103j3504P,Philcox:2021ukg} in 3D). $\mathcal{F}$ is estimated using 50 Monte Carlo realizations (using tricks analogous to \citep{2015arXiv150200635S}), and found to be extremely well converged \citep{Philcox:2023uwe}. Explicit definitions of each component can be found in \citep{Philcox:2023uwe}. This is a quasi-maximum-likelihood estimator, which, through the normalization matrix (which is approximately the variance of $\widehat{t}^{\rm num}$), accounts for leakage between bins and components of different parity, being unbiased for any choice of $\mathsf{S}^{-1}$.\footnote{Note that we assume the data and mask to be uncorrelated (since the signal is from recombination); see \citep{KristenPaper} for discussion of when this assumption breaks down and its potential amelioration.} The output spectra are close to minimum variance (with equality obtained in the limit of Gaussian maps and $\mathsf{S}^{-1}$ equal to the true inverse pixel covariance), can be \resub{(approximately)} compared to theory without mask convolution (cf.\,\citep{Hivon:2001jp} for the power spectrum), and can be efficiently computed (with $\mathcal{O}(N_{\rm bins})$ complexity). In limiting regimes, the inverse of the normalization matrix is equal to the analytic trispectrum covariance (\textit{i.e.}\ it is a Fisher matrix); this motivates working with the transformed variable
\beq\label{eq: tau-def}
    \widehat{\tau}(\vec b, B) \equiv \sum_{\vec b',B'}\mathcal{F}^{\rm{T}/2}(\vec b,B;\vec b',B')\widehat{t}(\vec b',B')
\eeq
(for Cholesky factorization $\mathcal{F}^{1/2}$ and transpose $\mathrm{T}$), which is close to a unit normal variable \citep[cf.,][]{Hamilton:2005ma}. \resub{Since $\F$ well describes the correlation structure of the data (validated in the Supplementary material), this accounts for correlations between bins (induced by the mask and cosmic variance), such that the individual $\tau$ bins are uncorrelated.}


Noting that the trispectrum dimensionality grows as $N_\ell^5$ (for $N_\ell$ bins per dimension), and the expected squared signal-to-noise scales as $\ell_{\rm max}^2$ (in the scale-invariant limit, cf.\,\citep{Kalaja:2020mkq}), we adopt a uniform binning in $\ell^{2/5}$, including $9$ bins in the range $[2,510]$,\footnote{The contiguous bins are defined by the edges $\{2, 3,  13,  35,  71, 121, 189, 275, 382, 510\}$. Notably, the signal-to-noise scaling depends on the model in question as discussed in \citep{Kalaja:2020mkq} and the Supplementary Material (including \citep{Kogo:2006kh,Bordin:2019tyb}), thus our binning may be suboptimal. For an analysis centered on a particular model type (e.g., a contact trispectrum), one would want to choose the binning accordingly, e.g., dropping all but collapsed tetrahedra.} 
dropping any configurations whose bin centers do not satisfy the triangle conditions. We fix $\ell_{\rm min}=2$ to avoid vanishing weights in \eqref{eq: weight-def}. \resub{Our choice} of $\ell_{\rm max}$ \resub{(and thus $N_{\rm side}$) is motivated both by computational considerations},\footnote{\resub{Theoretical computation of the trispectrum templates scales at least as $\ell_{\rm max}^3$, whilst, at fixed number of bins, the trispectrum computation scales with $N_{\rm side}^2$ (for each harmonic transform) and thus $\ell_{\rm max}^2$. Restricting to particular trispectrum configurations (e.g., collapsed or squeezed) can enable larger $\ell_{\rm max}$ without prohibitive computation times, though this goes against the spirit of a blind null test.}} and noting that parity-odd spectra are heavily suppressed at high-$\ell$ due to projection effects and Silk damping \citep{Kalaja:2020mkq}. To minimize leakage from unmeasured bins and foreground contamination, we simultaneously measure \resub{the numerators of} both parity-even and parity-odd \resub{trispectra} (a total of 1273 components, given the above bin restrictions), then \resub{discard the former (and the first bin) after the deconvolving normalization matrix has been applied \eqref{eq: polybin-eq}. This} leaving 460 parity-odd bins with $\ell_{\rm min}=3$, $\ell_{\rm max}=510$. Computation of the (data-independent) normalization matrix $\mathcal{F}$ requires $\approx 1000$ CPU-hours (and $200\,$GB memory, without optimization), with each simulation requiring an additional $10$ CPU-hours to analyze. In the Supplementary Material, we show the normalization matrices, $\mathcal{F}$, and present Fisher forecasts examining the impact of $\ell_{\rm max}$ on parameter constraints.

\section*{Blind Tests for Parity Violation}
\begin{figure}
    \centering
    \includegraphics[width=0.9\linewidth]{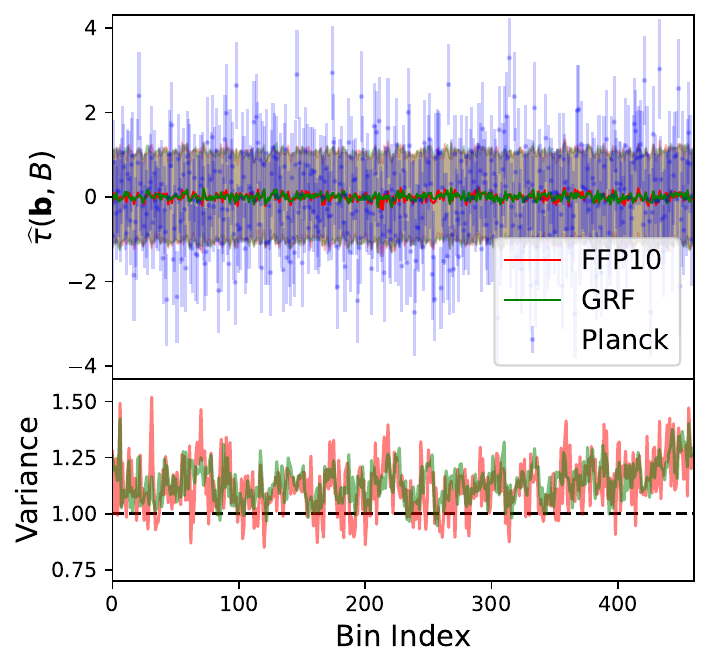}
    \caption{Normalized parity-odd trispectrum, $\tau(\vec b,B)$ measured from \textit{Planck} data (blue), 300 FFP10 simulations (red), and 1000 GRF simulations (green) using the \textsc{PolyBin} code with $\ell_{\rm min}=3$, $\ell_{\rm max}=510$. The horizontal axis gives the bin index (with large-scale modes on the left, ordered with $\ell_1\leq \ell_2$, $\ell_1\leq\ell_3$, $\ell_3\leq\ell_4$), and shaded regions show the $1\sigma$ variance from simulations (shown explicitly in the bottom panel). The variance is close to the Gaussian prediction (unity) but exhibits \resub{a slight enhancement} due to the non-trivial mask and finite number of bias simulations, though the correlation matrix is consistent with the identity (see the Supplementary material).}
    \label{fig: tau-data}
\end{figure}

\noindent To examine whether the \textit{Planck} data contains evidence for $\mathbb{P}$-violation, we first perform a blind model-agnostic test. Here, we follow a methodology similar to \citep{Philcox:2022hkh,Philcox:2021hbm,Cahn:2021ltp,Hou:2022wfj}, first computing the normalized trispectrum $\tau(\vec b,B)$ via \eqref{eq: tau-def}. This is shown in Fig.\,\ref{fig: tau-data}. For both the FFP10 and GRF simulations, we find a variance \resub{relatively} close to unity, indicating that our estimators are close to optimal \resub{(primarily within $\approx 20\%$, though with slightly larger deviations on small scales)}. \resub{Deviations} arise from (a) insufficient simulations used to remove disconnected components, (b) \resub{suboptimal weights $\mathsf{S}^{-1}$ (which could be rectified by a conjugate-gradient-inversion scheme)} (c) insufficient GRF simulations used to estimate the normalization matrix, (d) likelihood non-Gaussianity (but not large-scale foregrounds, as evidenced by the same behavior for the two sets of simulations). We have verified that (a) can be important (with larger variances seen when halving the number of bias simulations), but (c) is negligible, with a well-converged normalization matrix found even with five Monte Carlo realizations. \resub{Notably, suboptimal variances do not bias our analyses \citep[cf.,][]{Philcox:2023uwe}, though slightly degrade their constraining power.} In the Supplementary material, we verify that the correlation matrix does not show any noticeable departures from the identity. 

Data are analyzed via the following $\chi^2$ statistic:
\beq\label{eq: chi2}
    \widehat{\chi}^2 = \sum_{\vec b,B}\frac{\widehat{\tau}^2(\vec b, B)}{\mathrm{var}[\tau(\vec b,B)]},
\eeq
where the variance is computed from the GRF simulations, noting that the individual bins of $\tau$ are uncorrelated (which obviates the need for additional data compression). Under the following assumptions, the statistic can be analyzed using a $\chi^2$ distribution with 460 degrees of freedom: (1) $t$ obeys Gaussian statistics; (2) the GRF simulations accurately reproduce the \textit{Planck} trispectrum covariance. To test (1), we implement rank tests, comparing $\widehat{\chi}^2$ to the empirical $\chi^2$-distribution from the GRF (or FFP10) simulations. For (2), we note that the GRF simulations adopt a cosmology close to \textit{Planck}, and, unlike for LSS, higher-point functions entering the trispectrum covariance are negligible on the scales of interest. To test this assumption, we look for differences between the FFP10 and GRF results; we find excellent agreement, indicating that deviations of $\mathrm{var}(\tau)$ from unity arise primarily from masking and the finite number of bias realizations, which affect the simulations and data identically, \resub{\textit{i.e.}\ there is no evidence for contamination from residual foregrounds}. If the \textit{Planck} $\widehat{\chi}^2$ value lies significantly north of its expectation, we can conclude that there is evidence for large-scale $\mathbb{P}$-violation, since no conventional CMB physics would generate such a parity-odd signature. In contrast, a low $\widehat{\chi}^2$ value would imply systematic contamination.

\begin{figure}
    \centering
    \includegraphics[width=0.9\linewidth]{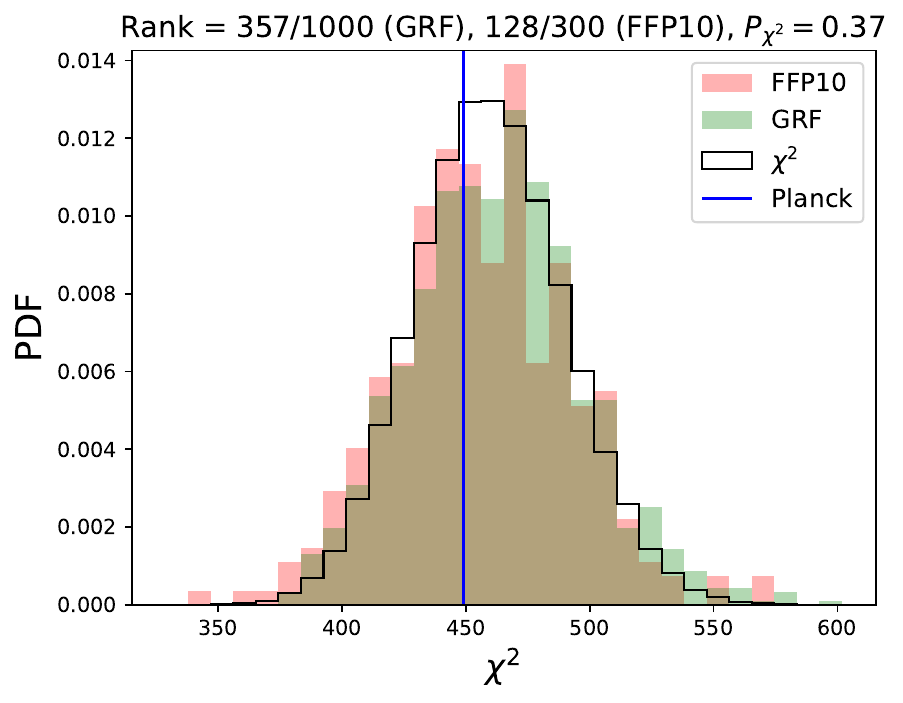}
    \caption{Probability of parity violation in the \textit{Planck} data-set. We plot the $\chi^2$ values \eqref{eq: chi2} extracted from the \textit{Planck} SMICA maps (blue), the FFP10 simulations (red), and GRF simulations (green), alongside the \resub{binned} theoretical prediction (black). Deviation of the \textit{Planck} results from the expected curves would give evidence for parity violation. Relative to the GRF simulations, the \textit{Planck} data has a rank of $357/1000$ (\textit{i.e.}\ a larger $\chi^2$ than that of $356$ simulations), or $128/300$ with respect to FFP10, with the consistency of the two sets of simulations indicating that our pipeline is robust \resub{and not affected by residual foregrounds}. For the theoretical distribution, we find a $\chi^2$ probability-to-exceed of $63\%$, though caution that the likelihood may be non-Gaussian. Overall, we find no evidence for parity violation in the \textit{Planck} dataset.}
    \label{fig: tau-results}
\end{figure}

Fig.\,\ref{fig: tau-results} shows the $\chi^2$ values extracted from the \textit{Planck} data, the 300 FFP10 simulations and the 1000 GRF realizations. Before discussing the data, we first note that the simulated distributions exhibit slightly \resub{wider posteriors} than the analytic prediction; this implies that the trispectrum likelihood is mildly non-Gaussian.\footnote{In initial testing using $\ell_{\rm max}=245$, but the same number of bins, this effect was significantly amplified.} If not accounted for (particularly in the distribution tails), this could lead to a false detection of parity violation. Secondly, we find similar empirical distributions from the simulation suites: relative to the GRF simulations, the FFP10 distribution has a mean probability-to-exceed (PTE) of $(46\pm2)\%$, relative to the expectation $(50\pm2)\%$, implying that any residual foreground-induced bias is small (though we note that the simulations do not include all correlations between foregrounds, which are relevant at large $\ell$ \citep[cf.,][]{Hill:2018ypf}).

We now consider the \textit{Planck} data. From the raw measurements shown in Fig.\,\ref{fig: tau-data} it is difficult to draw firm conclusions; however, Fig.\,\ref{fig: tau-results} allows us to compare the observed $\chi^2$ to the simulated distributions. Relative to the GRF suite, we find a rank of $357/1000$ (equivalent to a PTE of $64\%$ or $-0.4\sigma$) or $128/300$ (equivalent to $p=57\%$ or $-0.2\sigma$) with respect to FFP10. The conclusions are clear: \textbf{the large-scale \textit{Planck} temperature trispectrum shows no evidence for parity violation.}

\section*{Discussion}

\noindent In this \textit{Letter}, we have presented \resub{constraints on parity-violating physics from the large-scale \textit{Planck} temperature temperature, through a blind test for $\mathbb{P}$-violation and (in the Supplementary Material, which includes \citep{Arkani-Hamed:2003juy,Bucher:2015ura,2014A&A...571A..24P,Alonso:2018jzx,Foreman-Mackey:2012any,Bayer:2020pva,Cahn:2020axu,1988qtam.book.....V,Mehrem_1991,Niu:2022fki}), constraints on the amplitudes of various physical models}. We now consider the physical implications of \resub{the above}.

Previous constraints on scalar parity violation were obtained from LSS in \citep{Philcox:2022hkh,Hou:2022wfj}. To compare the two sets of results, it is useful to consider the number of linear modes present in each analysis. For the CMB, this is straightforward:
\beq
    N_{\rm modes}^{\rm CMB} = f_{\rm sky}\sum_{\ell=\ell_{\rm min}}^{\ell_{\rm max}}(2\ell+1)\left[\frac{C_\ell^{\rm TT}}{C_\ell^{\rm TT}+N_\ell^{\rm TT}}\right]^2,
\eeq
giving $\approx 2\times 10^5$ modes for $\ell_{\rm max}=510$ (with the signal-to-noise scaling roughly supported by the Fisher forecasts described in the Supplementary Material, implying that projection effects are subdominant on these scales). For galaxy surveys, we must account for the decorrelation of modes due to gravitational evolution. Using the approach of \citep{Sailer:2021yzm}, we can estimate the number of remaining linear modes at a fixed redshift $z$ as
\beq
    N_{\rm modes}^{\rm LSS}=V\int_{k_{\rm min}}^{k_{\rm max}}\frac{k^2dk}{2\pi^2}\frac{d\mu}{2}\left[\frac{G^2(k,\mu,z)P_{\rm lin}(k,\mu,z)}{P_{\rm tot}(k,\mu,z)}\right]^2
\eeq
where $V$ is the survey volume, $P_{\rm lin}$ and $P_{\rm tot}$ are the linear theory and total power spectra, and $G(k,\mu,z)\equiv \av{\delta_{\rm lin}\delta_{\rm tot}}/\av{\delta_{\rm lin}\delta_{\rm lin}}$ gives the decorrelation of modes from the initial conditions, estimated via a Zel'dovich propagator. Setting $k\in2\pi/[160,20]\hMpc$ and $V\approx 3.9h^{-3}\mathrm{Gpc}^{3}$, matching \citep{Philcox:2022hkh,Hou:2022wfj}, we find a total of $\approx 10^3$ linear modes, or $\approx 10^5$ if we ignore the propagator term, and assume all modes trace the initial conditions. 

The above comparison indicates that \textbf{the CMB dataset is a much stronger probe of the parity-violating initial conditions}. As such, our blind-test results suggest that the tentative detection of $\mathbb{P}$-violation seen in \citep{Philcox:2022hkh,Hou:2022wfj} is not primordial (since the $2.9\sigma$ hint of \citep{Philcox:2022hkh} would correspond to around $50\sigma$ in the CMB, based on primordial scaling arguments alone): instead, it could arise from exotic late-time physics (though see the caveats in \citep{Cabass:2022oap}), experimental artefacts, or analysis systematics (such as mocks of insufficient quality or likelihood non-Gaussianity). 

Turning to the constraints on physical models described in the Supplementary Material, we have found no evidence for any specific inflationary parity-breaking paradigm, with a maximal detection significance of $2.0\sigma$. Our limits on the Gauge model are broadly consistent with those forecasted in \citep{Shiraishi:2016mok} (at the level appropriate for the inexact nature of such predictions). Furthermore, the Fisher forecasts performed in the Supplementary Material yield similar constraints to the full analysis, implying that our posteriors are desirably Gaussian. Comparing to constraints from LSS, we find somewhat weaker bounds than those of former works \citep{Philcox:2022hkh,Cabass:2022oap}. This may appear to lie in contrast with the above mode-counting hierarchy; however, the LSS constraints did not account for mode decorrelation (instead assuming all modes were in the linear regime) or non-linear structure growth (which alters the scale-dependence of the signature), and did not marginalize over galaxy formation uncertainties. Each effect can lead to an inflation of the error-bars by several orders of magnitude.\footnote{The latter is demonstrated in \citep[Fig.\,3]{Cabass:2022epm}, whereupon $f_{\rm NL}$ constraints are inflated by $\sim$\,$100\times$ due to nuisance parameter marginalization even for the optimistic case of a high-redshift sample analyzed jointly with power spectrum data.} Furthermore, the precise constraints on physical models should be considered with caution since numerical inaccuracies in the template computation can lead to falsely enhanced precision (since noise in $\tau$ always increases the normalization matrix) -- the numerical convergence was more carefully considered for the templates used herein than for the LSS case.

Though the primordial templates generated by the three classes of models considered herein span a wide variety of the primordial space, they are not all-encompassing. Alternative models could include strong breaking of scale-invariance \citep{Cabass:2022rhr}, the exchange of higher spin particles, or the inclusion of inflationary modified gravity (e.g., with Chern-Simons interactions \citep{Alexander:2011hz,Alexander:2009tp,CyrilCS}). The bounds on such models can be straightforwardly computed given the relevant CMB templates, and proceeds analogously to the above analysis. An alternative option would be to devise a general basis for primordial parity-odd trispectra onto which any model can be projected, in analogy to the $f_{\rm NL}$ parametrization of bispectra.

Finally, we consider future prospects. On large-scales, the primary CMB is cosmic-variance-limited, such that the signal-to-noise on inflationary models will not increase with future experiments. However, one can consider the addition of polarization data (in particular $E$-modes), which can add substantial information on scalar physics \citep{Philcox:2023ypl}; in the case of $f_{\rm NL}^{\rm eq}$, for example, polarization provides $\approx 60\%$ of the constraining power \citep[Fig.\,1]{Planck:2019kim}. This will require modifying the trispectrum estimators of \citep{Philcox:2023uwe} to include fields of non-zero spin, but is not conceptually more complicated \citep{Philcox:2023psd}. Furthermore, one can extend the current analysis to smaller scales \resub{(for example making use of high-resolution ACT and SPT data)}, albeit at the cost of increased dimensionality and computation time. From the Fisher Forecasts performed in the supplementary material, doubling the $\ell$-range could increase the squared signal-to-noise by a factor of four (depending on the model in question), though we caution that projection effects start to limit analyses by $\ell_{\rm max}\sim 10^3$ \citep{Kalaja:2020mkq}. All in all, whilst these are the first constraints on CMB scalar parity violation, they seem unlikely to be the last.

\acknowledgments

\vskip 8 pt
{\footnotesize
\noindent We thank Stephon Alexander, William Coulton, Cyril Creque-Sarbinowski, Adriaan Duivenvoorden, Giulio Fabian, Colin Hill, Marc Kamionkowski, Daniel Meerburg, and William Underwood for insightful discussions that led to this work, \resub{as well as Neal Dalal, Krzysztof Gorski, Meng-Xiang Lin, Minh Nguyen, Ue-Li Pen and Maresuke Shiraishi for post-submission feedback}. We are additionally grateful to William Coulton, Cyril Creque-Sarbinowski, Colin Hill, and David Spergel for careful reading of the manuscript, \resub{as well as the anonymous referees for insightful feedback}. OHEP is a Junior Fellow of the Simons Society of Fellows and thanks John Watling of Nassau for liquid support. The author is pleased to acknowledge that the work reported in this paper was substantially performed using the Princeton Research Computing resources at Princeton University, which is a consortium of groups led by the Princeton Institute for Computational Science and Engineering (PICSciE) and the Office of Information Technology's Research Computing Division.
}
\bibliographystyle{apsrev4-1}
\bibliography{refs}


\newpage
\pagebreak

\appendix
\onecolumngrid

\pagebreak
\pagenumbering{arabic} 
\section{\Large Supplementary Material}

\section*{\texorpdfstring{I.\,\quad\,CONSTRAINTS ON PARITY-VIOLATING MODELS}{Constraints on Parity-Violating Models}}

\noindent The parity-odd trispectrum can be used to place constraints on various theories of primordial $\mathbb{P}$-violation. Here, we consider three classes of model, all arising from scale-invariant behavior during (approximately de-Sitter) inflation and specified by a parity-odd scalar curvature trispectrum:
\begin{enumerate}
    \item \textbf{Ghost}: Inflation with a Ghost condensate \citep{Arkani-Hamed:2003juy}, involving a quadratic dispersion relation. This contains two trispectra of interest, appearing at leading and subleading perturbative order. 
    \item \textbf{Collider}: Exchange of a massive spin-$1$ particle during inflation (possibly with broken boost symmetries, \textit{i.e.}\ varying sound speeds), following \citep{Cabass:2022rhr,Arkani-Hamed:2015bza}. 
    \item \textbf{Gauge}: Exchange of a $\mathrm{U}(1)$ gauge field with a Chern-Simons background and a non-vanishing vacuum expectation value, following \citep{Shiraishi:2016mok}. The Chern-Simons background leads to mode-function asymmetry and thus parity-violation at one-loop order. 
\end{enumerate}
The LSS imprints of these have been constrained using the BOSS data in \citep{Cabass:2022oap} (Ghost and Collider) and \citep{Philcox:2022hkh} (Gauge), which include further discussion of the models themselves.

To constrain such phenomena, we first require the relation between primordial and CMB trispectra. This is given in terms of the transfer function $\mathcal{T}_\ell(k)$ as
\beq\label{eq: Tl-T-curvature}
    \bigg\langle\prod_{i=1}^4a_{\ell_im_i}\bigg\rangle_c &=& (4\pi)^4\left[\prod_{i=1}^4i^{\ell_i}\int_{\vk_i}\mathcal{T}_{\ell_i}(k_i)Y^*_{\ell_im_i}(\hk_i)\right]\\\nonumber
    &&\times T_\zeta(\vk_1,\vk_2,\vk_3,\vk_4)\delD{\vk_{1234}},
\eeq
where $T_\zeta$ is the curvature trispectrum, $\vk_{1234}\equiv \sum_{i=1}^4\vk_i$, and the Dirac delta ensures momentum conservation. In \S\,\textcolor{red}{IV}, we sketch the derivation of each template given the primordial $T_\zeta$ shapes (which have been previously derived), yielding the final forms of \eqref{eq: ghost-tl},\,\eqref{eq: reduced-T-collider}\,\&\,\eqref{eq: gauge-tl}. Each model can be written $t^{\rm th} = A\,t^{\rm template}$, where the characteristic amplitudes $A$, are defined in \eqref{eq: A-ghost}\,\eqref{eq: A-collider}\,\&\,\eqref{eq: A-gauge} respectively. To compare to data, the templates are binned as follows (analogous to \citep{Bucher:2015ura,2014A&A...571A..24P} for the binned bispectrum):
\beq\label{eq: tl-theory}
    t^{\rm th}(\vec b,B) &\propto& \sum_{\ell_{1234}}\Theta_{\ell_1}(b_1)\Theta_{\ell_2}(b_2)\Theta_{\ell_3}(b_3)\Theta_{\ell_4}(b_4)\frac{(2\ell_1+1)(2\ell_2+1)(2\ell_3+1)(2\ell_4+1)(2L+1)}{(4\pi)^2}\\\nonumber
    &&\,\times\,\tj{\ell_1}{\ell_2}{L}{-1}{-1}{2}^2\tj{\ell_3}{\ell_4}{L}{-1}{-1}{2}^2\frac{t^{\ell_1\ell_2, \rm th}_{\ell_3\ell_4}(L)}{S_{\ell_1}S_{\ell_2}S_{\ell_3}S_{\ell_4}},
\eeq
where $S_\ell$ is the harmonic-space Wiener filter applied to the data. The normalization factor is of the same form but without the theoretical trispectrum. This weighting is derived by considering the expectation of the ideal trispectrum estimator, dropping the subdominant $6j$ term (which leads to a small mixing of $L$ modes).\footnote{\resub{Strictly, the weighting in each bin depends on the mask and $\mathsf{S}^{-1}$ function \citep[cf.,][]{Hivon:2001jp,Alonso:2018jzx}; such factors are heinously expensive to compute for trispectra, but drop out in the narrow bin limit, thus we assume the simplified form of \eqref{eq: tl-theory} in practice.}} Given the analytic results below, templates are computed numerically in \textsc{Python}, with the various integrals evaluated using numerical quadrature, following maximal factorization (as discussed below). These are computed using embarrassingly parallel multiprocessing (with Wigner symbols evaluated using \textsc{pywigxjpf}\footnote{\href{https://pypi.org/project/pywigxjpf/}{pypi.org/project/pywigxjpf}}), and require several tens of thousands of CPU-hours for each template (with appropriate choices of accuracy settings). This is the rate-limiting step of the analysis, due to the large number of coupled $\ell$ summations that must be evaluated, particularly for large $\ell_{\rm max}$. Our template generation code has been made publicly available online. 

\begin{figure}
    \centering
    \includegraphics[width=0.75\linewidth]{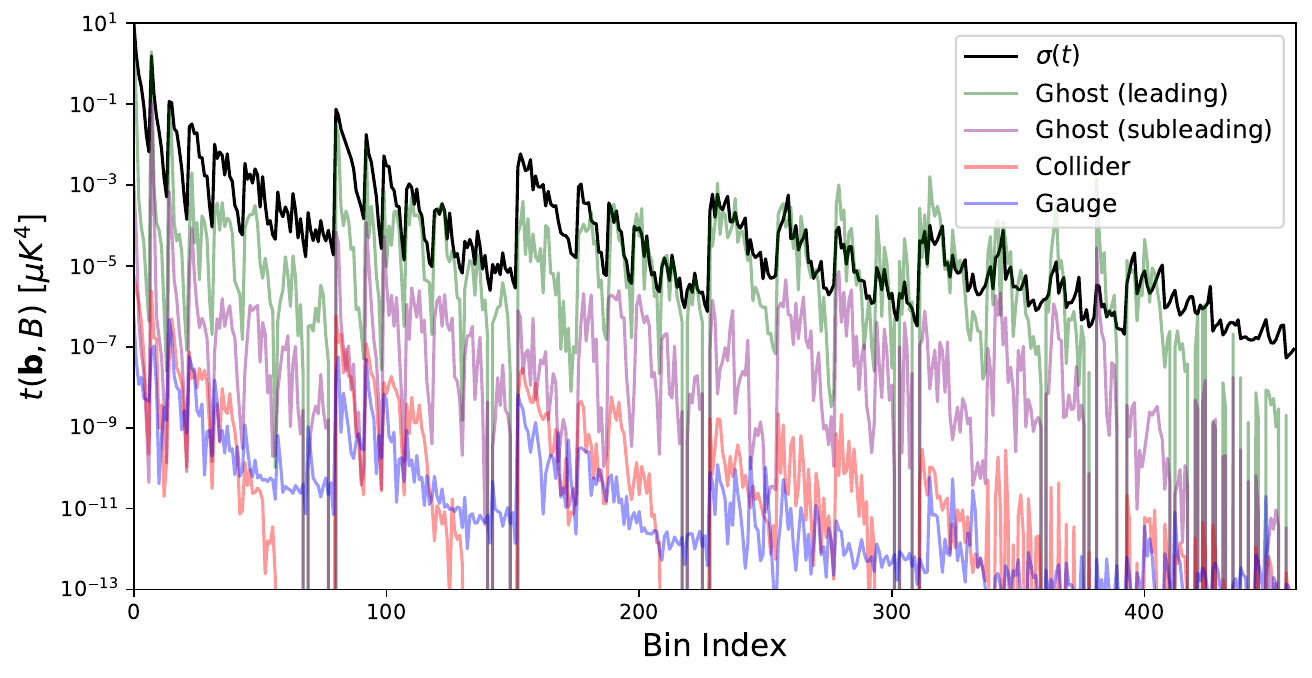}
    \caption{Theoretical templates for the parity-odd CMB trispectrum, plotted as a function of bin index. This includes modes from $\ell_{\rm min}=3$ (left) to $\ell_{\rm max}=510$ (right), and we additionally plot the standard deviation of the \textit{Planck} trispectrum trispectrum (black), as obtained from 300 FFP10 simulations (matching Fig.\,\ref{fig: var_all}). For the Collider model, we assume a mass of  $\nu=3/4$, and sound-speed $c_s=1$. For visualization, we have assumed fiducial amplitudes of $A^{(\rm Collider)} = 10^{-13}$, $A^{(\rm Ghost-I)} = 10^{-9}$, $A^{(\rm Ghost-II)} = 10^{-11}$, $A^{(\rm Gauge)} = 10^{3}$. In all cases, we find strong variation with the bin index, due to the scale-invariant nature of the primordial correlators.}
    \label{fig: tl-templates}
\end{figure}

Fig.\,\ref{fig: tl-templates} shows the numerical form of four such templates alongside the standard deviation of the \textit{Planck} trispectrum. In general, these plots are difficult to interpret, which is a natural consequence of attempting to visualize high-dimensional data. However, we can observe strong trends in $\ell$ for each model (from left to right): these arise from the assumption of scale invariance in the inflationary correlators (\textit{i.e.}\ a de Sitter background background). Secondly, we find qualitatively different behavior between templates, due to their different microphysical formulations, in particular the presence or absence of a coupled inflationary field. On small scales, most templates are suppressed compared to the variance; this is expected, since we approach the flat-sky regime where parity-transformations and rotations are equivalent.

\begin{table}[htb]
    \centering
    \begin{tabular}{l|c|c}
    \textbf{Model} & \textit{Planck} & FFP10\\\hline
    Ghost (leading, $\times 10^{12}$) &  $-0.7 \pm 1.4$ & $0.3 \pm 1.4$\\
    Ghost (subleading, $\times 10^{13}$) &  $4.0 \pm 7.5$ & $-1.4 \pm 7.5$\\
    Collider ($\nu=3/4,c_s=1, \times 10^{10})$ & $-2.6 \pm 4.3$ & $-0.3 \pm 4.3$\\
    Collider ($\nu=0,c_s=1, \times 10^{10})$ & $-5.5 \pm 7.3$ & $-0.6 \pm 7.4$\\
    Collider ($\nu=5/4,c_s=1, \times 10^{10})$ & $-0.8 \pm 1.6$ & $-0.1 \pm 1.6$\\
    Collider ($\nu=3/4,c_s=0.1, \times 10^{14})$ & $0.4 \pm 5.0$ & $0.0 \pm 5.0$\\
    Collider ($\nu=3/4,c_s=10, \times 10^4)$ & $-1.0 \pm 3.1$ & $-0.2 \pm 3.1$\\
    Gauge ($\times 10^{-7}$) & $-2.8 \pm 1.3$
     & $-0.1 \pm 1.4$\\
    \end{tabular}
    \caption{Constraints on physical models of parity violation from the \textit{Planck} temperature trispectrum. In each case, we give a $1\sigma$ constraint on the relevant physical amplitude $A$, whose form is given in the Supplementary Material. For the Collider scenario, we test five representative parameter sets, including the conformally coupled scenario ($\nu = 0$). Results are additionally shown from the mean of 300 FFP10 simulations, and indicate that our pipeline is unbiased. We find no significant detection of any primordial template, with a maximal deviation of $2.0\sigma$ (\textit{Planck}, Gauge) or $0.2\sigma$ (FFP10, Ghost leading).}
    \label{tab: model-constraints}
\end{table}

\begin{figure}
    \centering
    \includegraphics[width=0.9\textwidth]{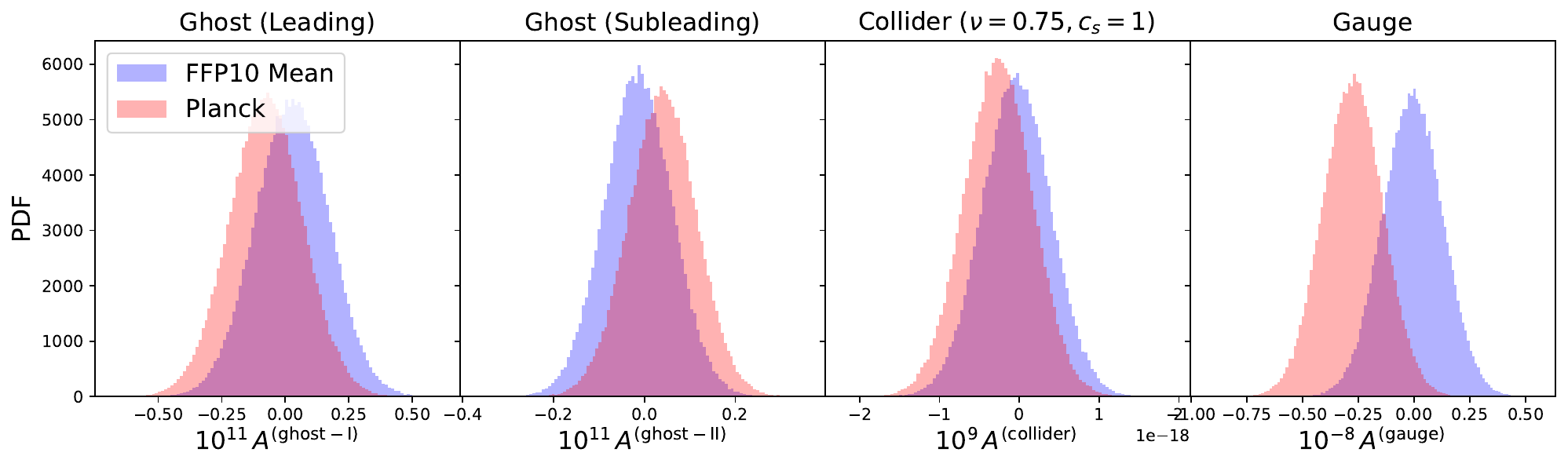}
    \caption{Constraints on the amplitudes of physical models for parity violation using the \textit{Planck} dataset (red) and the mean of 300 FFP10 simulations (blue). Here, we show results corresponding to the four templates shown in Fig.\,\ref{fig: tl-templates}, whose limits are tabulated in Tab.\,\ref{tab: model-constraints}. We find no evidence for any model of primordial parity violation, with a maximal detection significance of $2.0\sigma$ ($0.2\sigma$) across eight models for \textit{Planck} (FFP10) data.}
    \label{fig: model-results}
\end{figure}

To constrain the inflationary models, we fit for $A$ by first projecting onto the normalized basis with $\tau^{\rm template}\equiv \mathcal{F}^{\rm T/2}t^{\rm template}$ (involving only parity-odd modes), then sampling the likelihood
\beq
    \mathcal{L}(A) \propto \mathrm{exp}\left(-\sum_{\vec b,B}\frac{\left[\widehat{\tau}(\vec b,B) -A\,\tau^{\rm th}(\vec b,B) \right]^2}{\mathrm{var}\left[\tau(\vec b,B)\right]}\right),
\eeq
using \textsc{emcee} \citep{Foreman-Mackey:2012any}.\footnote{We may use a Gaussian likelihood for $A$ since we have projected the trispectrum onto a single degree of freedom, thus the central limit theorem can be applied.} The resulting constraints are given in Tab.\,\ref{tab: model-constraints}, with representative posteriors shown in Fig.\,\ref{fig: model-results}. Results from the FFP10 simulation suite show that our pipeline is unbiased: all constraints are consistent with zero within $0.2\sigma$. Although the previous section already indicated that the simulations showed no evidence for $\mathbb{P}$-violation, this is a more stringent test, since we are optimally projecting onto the template interest. \textbf{From the \textit{Planck} data, we find no evidence of any inflationary model}. The maximum detection significance is $2.0\sigma$ (for the Gauge model), and we note that the chances of finding outliers is enhanced since we perform eight different analyses \citep{Bayer:2020pva}. For the Collider templates, we find significant variation in constraining power with the microphysical parameters $c_s$ and $\nu$: in particular, models with smaller sound-speeds are more strongly constrained, matching the conclusions of \citep{Cabass:2022rhr}. In the main paper discussion, we compare these results to other constraints found in the literature.

\section{\texorpdfstring{II.\quad\,NORMALIZATION MATRICES \& COVARIANCES}{Normalization Matrices \& Covariances}}\label{subsec: norm-matrices}

\begin{figure}
    \centering
    \includegraphics[width=0.6\linewidth]{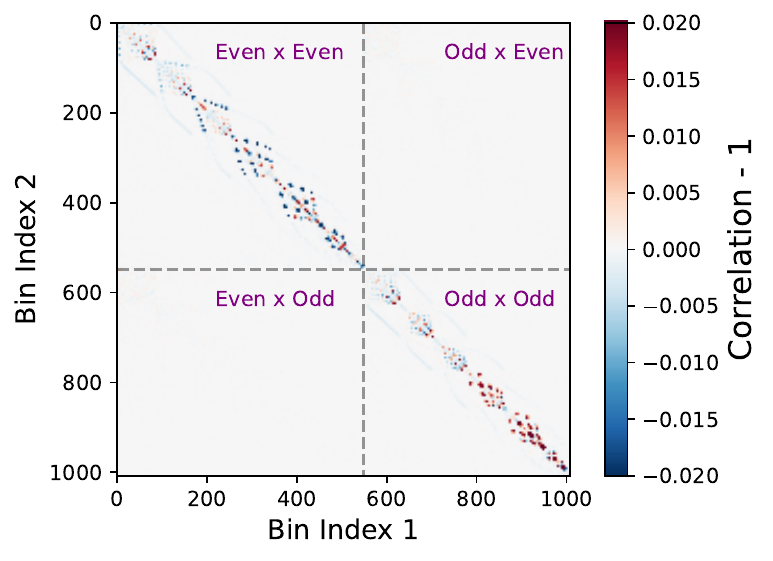}
    \caption{Normalization matrix for the parity-even and parity-odd trispectrum, as measured by the \textsc{PolyBin} code using 50 Monte Carlo realizations. This is used to deconvolve the mask and Wiener filter from the raw trispectrum measurements. For visualization, we normalize by the diagonal elements (themselves shown in Fig.\,\ref{fig: var_all}), and choose the scale so as to emphasize small contributions. The complex non-diagonal structure arise from bin-to-bin correlations induced by the mask; these are primarily concentrated in adjacent trispectrum bins (which are spread out in the plot due to bin ordering). We additionally find little correlation between parity-even and parity-odd modes.}
    \label{fig: fisher}
\end{figure}

\noindent In this section, we present and validate the normalization matrix, $\mathcal{F}$, computed by the \textsc{PolyBin} code. As described above, this is used to remove bin-to-bin correlations imprinted by the mask, and additionally reduces leakage between parity-even and parity-odd modes. Fig.\,\ref{fig: fisher} displays the correlation structure of the matrix including both parities. The aforementioned correlations yield a complex (almost artistic) form, with largest signals (up to $\approx 10\%$) found close to the diagonal. These arise from the window-function-induced leakage between neighboring trispectrum bins.\footnote{Note that the \textsc{PolyBin} algorithm gives an unbiased estimate of the trispectrum, $t$; the projection matrix in \eqref{eq: tau-def} then imprints mask information onto $\tau$, but this does not yield bias, since the same transformation is applied to any theoretical model, and mean-zero $t$ implies mean-zero $\tau$. Some deviation of the variance from unity is expected for complex masks; the full covariance is given by $\mathcal{F}^{-\rm T/2}[\mathsf{S}^{-1}]\mathcal{F}[\mathsf{S}^{-1}\mathsf{C}\,\mathsf{S}^{-\rm T}]\mathcal{F}^{-1/2}[\mathsf{S}^{-1}]$, where $\mathcal{F}[\mathsf{M}]$ is the normalization matrix with weighting $\mathsf{M}$ and $\mathsf{C}$ is the true pixel covariance.} Notably, we find little correlation between modes of different parities, with only a slight hint seen for the lowest bins (\textit{i.e.}\ largest scales).

\begin{figure}
    \centering
    \includegraphics[width=0.8\linewidth]{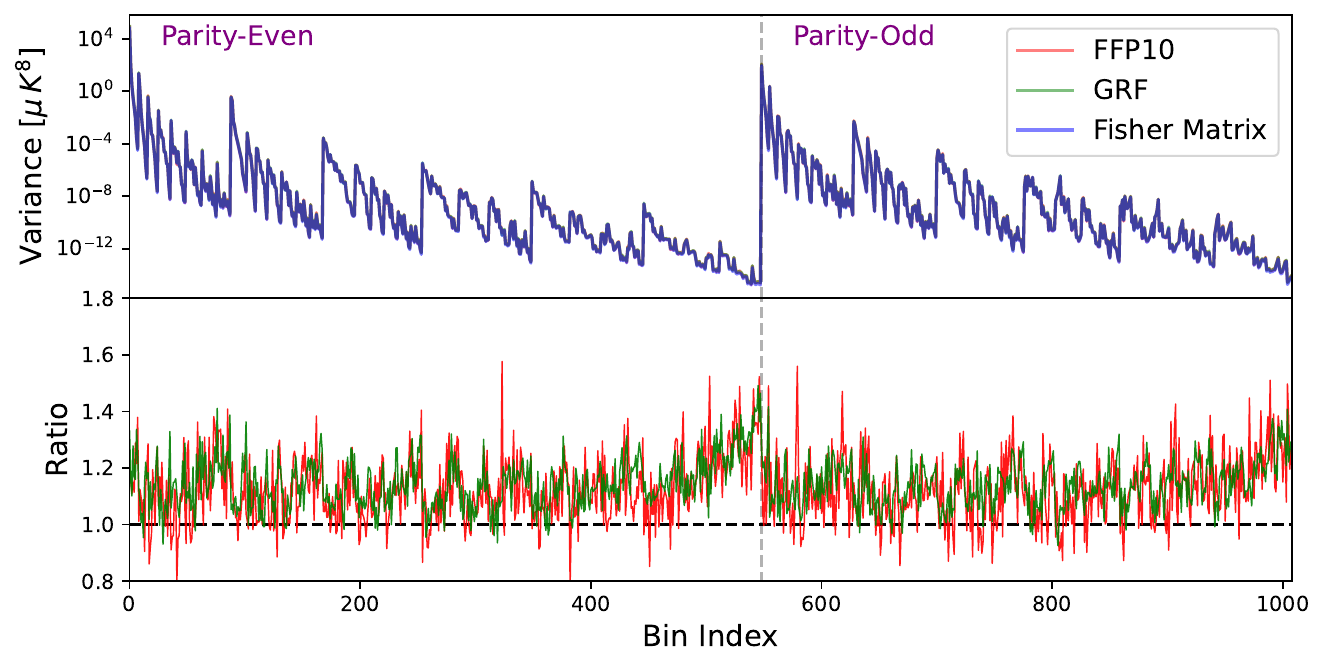}
    \caption{Comparison of the inverse normalization matrix (blue; Fisher) and the trispectrum variances obtained from the FFP10 (red) and GRF simulations (green). We display results for both parity-even (left) and parity-odd (right) trispectra, with the bottom panel showing the ratios with respect to the normalization matrix. The sawtooth pattern is not a physical signature, but an artefact of the compression of five-dimensional data to a one-dimensional axis, appearing due to the ordering of $\ell_i$ and $L$ \citep[cf.,][]{Philcox:2023uwe,Philcox:2021ukg}. If the estimator is optimal, the inverse normalization matrix should be equal to the trispectrum variance; here we find that this holds to within \resub{$\approx 20\%$}, and additionally find good agreement between the two simulation suites.}
    \label{fig: var_all}
\end{figure}

In Fig.\,\ref{fig: var_all}, we plot the diagonal elements of the inverse normalization matrix alongside measurements of the empirical trispectrum variance from the FFP10 and GRF simulation suites (without rescaling by $\mathcal{F}^{\rm T/2}$). If the estimator is optimal, the two should match; here, we find good agreement across seventeen orders of magnitude and for both parities. Considering the ratios, we find that the numerical variances are, on average, around $20\%$ higher than the optimal limit, and consistent between the two sets of simulations. This indicates that (a) the estimator is close to optimal (with a residual $\approx 10\%$ inflation in error-bars, \resub{as found in the main text}), and (b) any residual non-Gaussian signatures in the FFP10 simulations are small.

\begin{figure}
    \centering
    \includegraphics[width=0.5\linewidth]{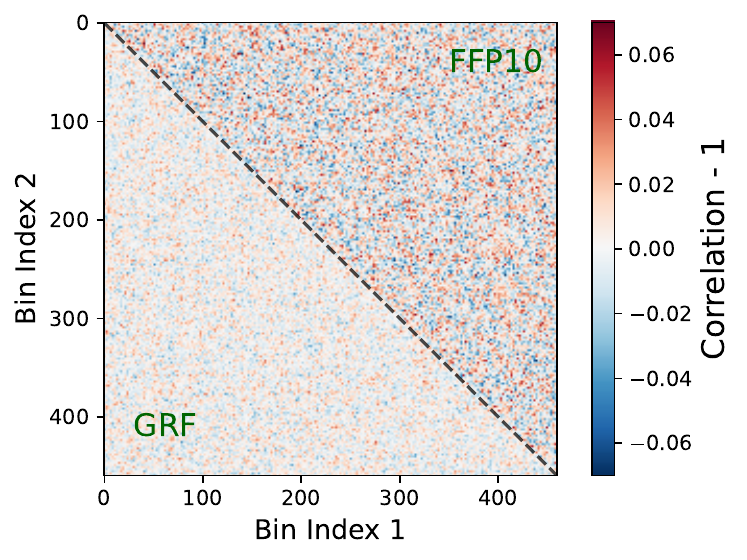}
    \caption{Correlation matrix of the normalized parity-odd trispectrum shown in Fig.\,\ref{fig: tau-data}. The upper right and lower left regions show the correlation matrix for the FFP10 and GRF simulations respectively. We subtract off a unit matrix for clarity, such that the mean correlation is expected to be zero if the trispectrum bins are uncorrelated.}
    \label{fig: tau-corr}
\end{figure}

To fully validate the normalization matrix $\mathcal{F}$ we must also look at the off-diagonal components. This is easiest to perform using the rescaled statistic $\tau$: if the estimator is optimal, the correlation matrix of $\tau$ (equal to the covariance, normalized by its diagonal), should be a unit matrix. In Fig.\,\ref{fig: tau-corr}, we show this for the two sets of simulations. In both cases, we find no discernible departures from the identity, again demonstrating that our estimators are close to optimal. This additionally motivates the assumption of independent $\tau$ components when building our $\chi^2$ statistic.

\section{\texorpdfstring{III.\quad\,FISHER FORECASTS}{Fisher Forecasts}}
\noindent To examine the dependence of parameter constraints on our choice of scale cuts, we perform a series of idealized Fisher forecasts. To this end, we first compute the trispectrum normalization matrix, $\mathcal{F}$, using the \textsc{PolyBin} code, working in the simplified limit of a unit mask with no coupling between trispectrum bins. Under ideal assumptions, this is equal to the inverse covariance of the parity-odd trispectrum (see \citep{Philcox:2023uwe} for the explicit forms). Given a trispectrum template $A\,t^{\rm model}$, we forecast the $1\sigma$ error on its amplitude $A$ via
\beq\label{eq: fisher-forecast}
    \sigma_{A}^{-2} \leq \sum_{\vec b,\vec b',B,B'}t^{\rm model}(\vec b,B)\mathcal{F}(\vec b,B;\vec b',B')t^{\rm model}(\vec b',B'),
\eeq
for bins $\vec b,B$, which simplifies given that the above normalization matrix is diagonal. 

For the three classes of models discussed herein, we forecast the constraining power on the relevant amplitudes in three regimes: (a) `full', using all bins up to a given $\ell_{\rm max}$; (b) `collapsed', restricting to low $L_{\rm max}$ for the internal momentum; (c) `squeezed', fixing $\ell_1$ and $\ell_3$ to low values, but allowing other multipoles to extend to $\ell_{\rm max}$. The latter two options test the dependence on quadrilateral configurations, which we expect to differ depending on the physical model in question.
In all cases, we compute the raw spectra assuming the same binning as in the main text (with $\ell_{\rm min}=3$, $\ell_{\rm max}=510$), and additionally fix $\nu=3/4$, $c_s=1$ for the Cosmological Collider templates. As in the main text, we consider only bins whose centers satisfy the relevant triangle conditions; we caution that this can lead to some loss of signal-to-noise at high-$\ell$ where the bins are wide.

\begin{figure}
    \centering
    \includegraphics[width=0.8\textwidth]{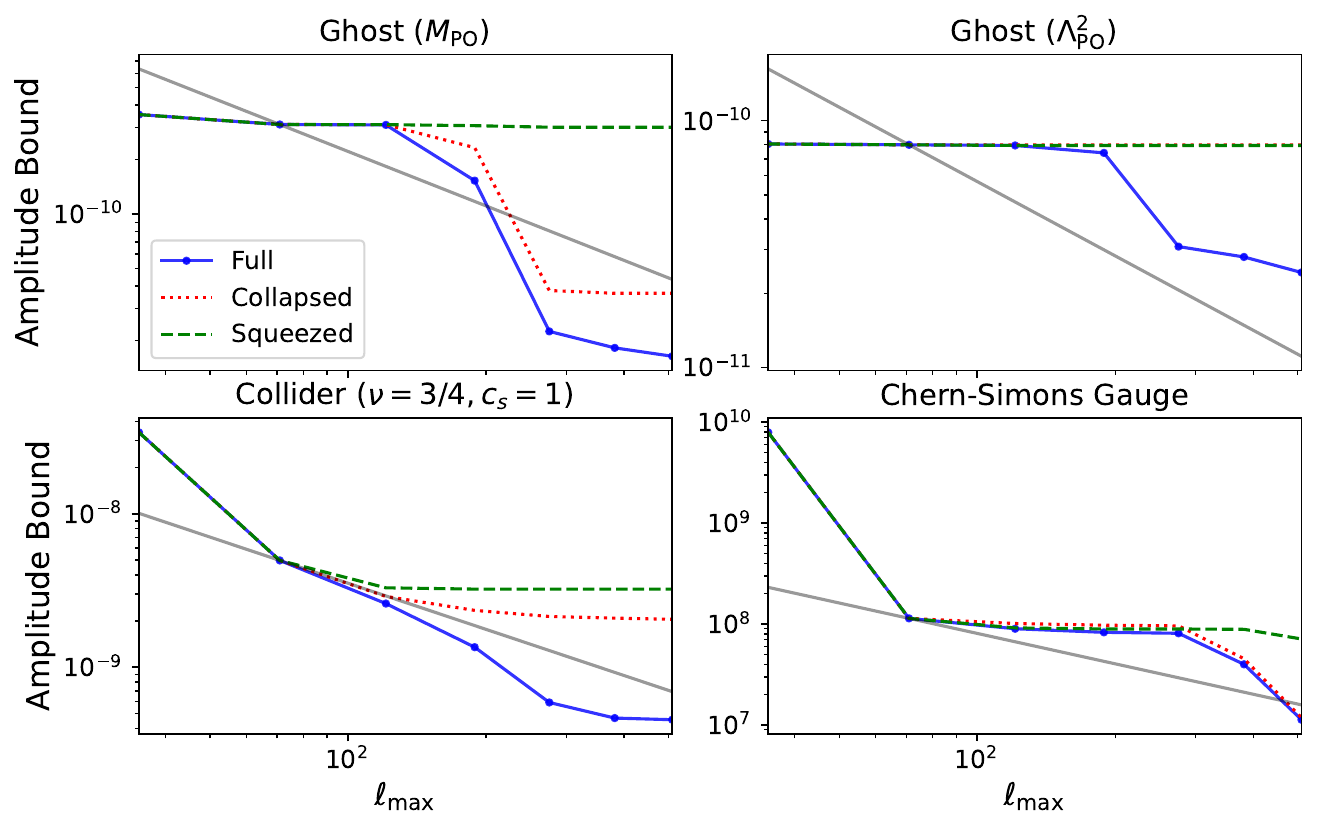}
    \caption{Forecasted constraints on the amplitudes of four inflationary trispectrum models as a function of the maximum scale used in the analysis. We assume an idealized scenario without an observational mask or bin correlations, and perform a Fisher forecast, as in \eqref{eq: fisher-forecast}. The full forecasts (using subsets of the total trispectrum) are shown in blue, whilst the red curves shows forecasts with $L_{\rm max}\approx 100$, and the green shows the impact of allowing for squeezed triangles up to $\ell_{\rm max}$ but restricting unsqueezed legs to $\ell\lesssim 100$.
    The gray lines show the idealized $1/\ell_{\rm max}$ scalings. We find that the scaling with $\ell_{\rm max}$ is strongly dependent on the type of primordial interactions: exchange models (Collider and Gauge) have a signal-to-noise increasing as $\ell_{\rm max}$, but contact trispectra (Ghosts) contain most information in the lowest $\ell$-bins.}
    \label{fig: fisher-forecast}
\end{figure}

Before discussing the numerical results, let us first outline our expectations. According to mode-counting arguments, the squared signal-to-noise of a scale-invariant primordial trispectrum should scale as $\ell_{\rm max}^2$, 
motivating the binning scheme used in the main text. In practice, however, the situation is less straightforward. At large scales, diffusion damping washes out information (leading to a $\log\ell_{\rm max}$ scaling at large $\ell$ \citep{Kalaja:2020mkq}), and for parity-odd spectra, projection effects lead to a loss of information, since any violations are washed out in the flat-sky limit by rotational invariance. In our contexts, we expect the latter effect to be dominant, since we restrict ourselves to comparatively large scales, where Silk damping is small. Furthermore, as demonstrated in \citep{Kogo:2006kh,Bordin:2019tyb,Kalaja:2020mkq}, the signal-to-noise of a given model depends heavily on its shape, \textit{i.e.}\ whether it peaks in equilateral, squeezed, or collapsed configurations. For example, a primordial template of the exchange form (e.g., the Collider template used below) with a mass of $\nu\in(0,3/2)$ has a squared signal-to-noise scaling of $\ell_{\rm max}^{2}$ in the squeezed limit but finds a slower scaling in the collapsed limit if $\nu\geq 3/4$.

Fig.\,\ref{fig: fisher-forecast} shows the forecasted amplitude constraints (or inverse signal-to-noise, at $A=1$) from our pipeline, compared to the na\"ive $\sigma(A)\propto \ell_{\rm max}^{-1}$ scaling. For the Collider and Gauge scenarios, we find that the expected scaling is generally realized, with a gradual increase in signal-to-noise at larger $\ell_{\rm max}$. In contrast, $\sigma(A)$ is roughly flat with $\ell_{\rm max}$ for the Ghost models at low-$\ell$ (but we find improvement at larger $\ell$, though we caution that this is where numerical errors may lurk): this implies that the majority of primordial information is concentrated in the first few bins. Both sets of results match expectations, since the Collider and Gauge scenarios involve exchange diagrams, which peak at low $L$, but are less sensitive to the other legs, whilst the Ghost diagrams are of the contact form, whose diagrams dominate when all $\ell_i$ legs are small. Considering the tetrahedral configurations themselves, we find that the Ghost model loses significant signal-to-noise if we restrict to squeezed and (to a lesser extent) collapsed configurations: this occurs due to the resulting reduction in the number of modes. For the contact diagrams, restricting to collapsed tetrahedra (at low $L$) has a lesser impact on the signal-to-noise, particularly for the Gauge model; this supports our earlier statement and matches the findings of \citep{Shiraishi:2016mok}. Significant loss of information is found when restricting to squeezed triangles, which is again caused by the reduction in the number of primordial modes. 

From these forecasts, we may conclude that our results have important dependence on scale cuts. If our focus was limited to primordial trispectra of the contact form (e.g., the Gauge model), we may restrict to lower $L$, which would significantly expedite computation; however, such limits are model-specific. Pushing the analysis to larger $\ell_{\rm max}$ is likely to yield improved constraints on various models (with a rough $\sigma(A)\propto \ell_{\rm max}^{-1}$ scaling found), though the constraints on contact diagrams may improve more moderately. We finally note that our results will additionally depend on the width of our $\ell$-bins: broad bins lose signal at large $\ell$, and are less powerful at model discrimination: however, using finer bins comes at the cost of increased data computation time (which scales as $\mathcal{O}(N_{\rm bins})$), and a more non-Gaussian likelihood.

\section{\texorpdfstring{IV.\quad\,TEMPLATE DERIVATIONS}{Template Derivations}}\label{sec: derivs}

\noindent To derive the CMB templates, we will adopt the following procedure, allowing us to simplify the various components of \eqref{eq: Tl-T-curvature}: (1) compute the inflationary trispectrum $T_\zeta$, before symmetrization over $\vk_i$, (2) rewrite the Dirac delta functions as radial and angular pieces, (3) decompose any angular pieces in a spherical harmonic basis, (4) integrate over all spherical harmonics. This will yield a form involving only coupled one-dimensional integrals and angular momentum summations. Notably, the theoretical models are much simpler for the CMB trispectra than for LSS \citep{Cabass:2022oap}; this occurs since the latter works in a non-trivial angular basis involving an extra set of spherical harmonics. The resulting spectra are plotted in Fig.\,\ref{fig: tl-templates}.

\subsection{Ghost Inflation}\label{subsec: ghost-calc}
We begin with the scalar trispectrum induced by ghost inflation, as defined in \citep{Cabass:2022rhr,Cabass:2022oap} at leading and subleading order:
\beq\label{eq: ghost-Tk}
    \tilde{T}_{M_{\rm PO}}(\bfk_1,\bfk_2,\bfk_3,\bfk_4) &=& {\frac{128i\pi^3 \Lambda^5(H\tilde{\Lambda})^{1/2}}{M_{\rm PO} \tilde{\Lambda}^5 \Gamma(\frac{3}{4})^2}}(\Delta^2_\zeta)^3\frac{(\k_1\cdot\k_2\times \k_3)(\k_2\cdot \k_4)(\k_1\cdot \k_4)(\k_2\cdot \k_3)}{k_1^{\frac{3}{2}}k_2^{\frac{3}{2}}k_3^{\frac{3}{2}}k_4^{\frac{3}{2}}}\,{\rm Im}\,{\cal T}^{(11)}_{0,0,0,0}(k_1,k_2,k_3,k_4)\nonumber\\ 
    \tilde{T}_{\Lambda^2_{\rm PO}}(\bfk_1,\bfk_2,\bfk_3,\bfk_4) &=& {\frac{512i\pi^3 \Lambda^5(H\tilde{\Lambda})^{3/2}}{\Lambda_{\rm PO}^2 \tilde{\Lambda}^6 \Gamma(\frac{3}{4})^2}}(\Delta^2_\zeta)^3(\k_1\cdot\k_2\times \k_3)(\k_1\cdot \k_2) k_1^{-\frac{3}{2}}k_2^{\frac{1}{2}}k_3^{\frac{1}{2}}k_4^{\frac{1}{2}}\,{\cal T}^{(13)}_{0,0,0,1}(k_1,k_2,k_3,k_4), 
\eeq
where $H$ is the Hubble parameter during (de-Sitter) inflation, $\Delta_\zeta^2$ is the power spectrum amplitude, and $\Lambda^2_{\rm PO}$, $\tilde{\Lambda}$, $M_{\rm PO}$ describe the microphysical properties of the model, setting the amplitude of parity violation. We additionally write the trispectrum in pre-symmetrized form as $T(\vk_1,\vk_2,\vk_3,\vk_4)\equiv\delD{\vk_{1234}}\tilde{T}(\vk_1,\vk_2,\vk_3,\vk_4)+(\text{23 perms})$. Here, $\mathcal{T}^{(n)}$ is given by
\beq\label{eq: ghostT}
    \smash{{\cal T}^{(n)}_{\nu_1,\nu_2,\nu_3,\nu_4}(k_1,k_2,k_3,k_4)}  = \int_{0}^{+\infty}{\rm d}\lambda\, \lambda^{n}\,H^{(1)}_{{\frac{3}{4}-\nu_1}}(2ik^2_1\lambda^2)H^{(1)}_{{\frac{3}{4}-\nu_2}}(2ik^2_2\lambda^2)H^{(1)}_{{\frac{3}{4}-\nu_3}}(2ik^2_3\lambda^2)H^{(1)}_{{\frac{3}{4}-\nu_4}}(2ik^2_4\lambda^2). 
\eeq 
For ghost inflation, the trispectrum is of the contact form, since there is no exchange momentum. This model will be constrained via the amplitudes
\beq\label{eq: A-ghost}
    A^{(\rm Ghost -I)} = \frac{\Lambda^5(H\tilde{\Lambda})^{1/2}}{M_{\rm PO} \tilde{\Lambda}^5 \Gamma(\frac{3}{4})^2}(\Delta^2_\zeta)^3, \qquad A^{(\rm Ghost -II)} = \frac{\Lambda^5(H\tilde{\Lambda})^{3/2}}{\Lambda_{\rm PO}^2 \tilde{\Lambda}^6 \Gamma(\frac{3}{4})^2}(\Delta^2_\zeta)^3
\eeq
matching \citep{Cabass:2022oap} (though relabelled).

To proceed, we expand the Dirac delta function in terms of spherical harmonics. This takes the form \citep[Eq.\,A7][]{Cabass:2022oap}
\beq
    \delD{\vk_{1234}} &=& \int {\rm d}\vx\,e^{i\vk_{1234}\cdot\vx} \nonumber \\ \nonumber
    &=&(4\pi)^4\sum_{L_1\cdots L_4M_1\cdots M_4}i^{L_{1234}}\int x^2 {\rm d}x\int {\rm d}\hx\,\left[\prod_{i=1}^4 \sum_{M_i}j_{L_i}(k_ix)Y_{L_iM_i}(\hk_i)Y^*_{L_iM_i}(\hx)\right] \\ 
    &=&(4\pi)^3\sum_{L_1\cdots L_4}(-i)^{L_{1234}}\int x^2 dx\,j_{L_1}(k_1x)j_{L_2}(k_2x)j_{L_3}(k_3x)j_{L_4}(k_4x)\\\nonumber
	&&\,\times\,\sum_{L'}\tjo{L_1}{L_2}{L'}\tjo{L'}{L_3}{L_4}\mathcal{C}_{L_1L_2L_3L_4L'}\P_{L_1L_2(L')L_3L_4}(\hk_1,\hk_2,\hk_3,\hk_4),
\eeq
where we perform Gaunt integrals to obtain the final equality. This involves the function $\mathcal{C}_{\ell_1\cdots \ell_n}=\sqrt{(2\ell_1+1)\cdots(2\ell_n+1)}$, as well as an isotropic basis function of four $\hk$ variables, $\mathcal{P}$, as defined in \citep{Cahn:2020axu} (or equivalently a polypolar spherical harmonic \citep[e.g.,][]{1988qtam.book.....V}). Next, the angular components of \eqref{eq: ghost-Tk} can be written as a combination of isotropic basis functions \citep[Eq.\,A10-11]{Cabass:2022oap}
\beq
    \left(\hk_1\cdot\hk_2\times\hk_3\right)(\hk_2\cdot\hk_4)(\hk_1\cdot\hk_4)(\hk_2\cdot\hk_3) &=& -i\frac{\sqrt{2}}{9\sqrt{3}}(4\pi)^{8}\left[\P_{11(1)10}\P_{01(1)01}\P_{10(1)01}\P_{01(1)10}\right](\hk_1,\hk_2,\hk_3,\hk_4)\nonumber\\\nonumber
    &=&-i\frac{\sqrt{10}}{225}(4\pi)^2\left[\P_{01(1)22}-2\P_{03(3)22}+\frac{4\sqrt{2}}{5}\P_{21(1)22}+\P_{21(2)20}\right.\\
    &&\qquad-\sqrt{\frac{14}{5}}\P_{21(2)22}+\frac{2\sqrt{7}}{5}\P_{21(3)22}+\frac{\sqrt{3}}{5}\P_{23(1)22}-2\P_{23(2)20}\\\nonumber
    &&\qquad\left.-\sqrt{\frac{2}{35}}\P_{23(2)22}-\frac{2\sqrt{3}}{5}\P_{23(3)22}+\frac{6}{\sqrt{7}}\P_{23(4)22}\right](\hk_1,\hk_2,\hk_3,\hk_4)\\\nonumber
    &\equiv&-i(4\pi)^2\sum_{l_1l_2l_3l_4l'}c_{l_1l_2(l')l_3l_4}\P_{l_1l_2(l')l_3l_4}(\hk_1,\hk_2,\hk_3,\hk_4),\\\nonumber
    \left(\hk_1\cdot\hk_2\times\hk_3\right)(\hk_1\cdot\hk_2) &=& -i\frac{\sqrt{2}}{3\sqrt{3}}(4\pi)^{3}\P_{111}(\hk_1,\hk_2,\hk_3)\P_{110}(\hk_1,\hk_2,\hk_3) \\
	&=& -\frac{i}{3}\sqrt\frac{2}{5}(4\pi)^2\P_{22(1)10}(\hk_1,\hk_2,\hk_3,\hk_4),
\eeq
where we define coefficients $c_{l_1l_2(l')l_3l_4}$ for brevity.

Combining terms, the $M_{\rm PO}$ and $\Lambda_{\rm PO}^2$ templates become, respectively,
\beq\label{eq: ghost-Tl}
    T^{\ell_1\ell_2\ell_3\ell_4}_{m_1m_2m_3m_4} &=& 2(4\pi)^8 A^{(\rm Ghost -I)}\sum_{l_1l_2l_3l_4l'}c_{l_1l_2(l')l_3l_4}\sum_{L_1\cdots L_4L'}i^{\ell_{1234}-L_{1234}}\tjo{L_1}{L_2}{L'}\tjo{L'}{L_3}{L_4}\mathcal{C}_{L_1L_2L_3L_4L'}\nonumber\\
    &&\,\times\,{\rm Im}\int_{0}^{\infty}\lambda^{11}d\lambda\,\int x^2 dx\,h^{\ell_1L_1}_{0,1/2}(x,\lambda)h^{\ell_2L_2}_{0,3/2}(x,\lambda)h^{\ell_3L_3}_{0,1/2}(x,\lambda)h^{\ell_4L_4}_{0,1/2}(x,\lambda)\\\nonumber
	&&\,\times\,\prod_{i=1}^4\left[\int d\hk_iY^*_{\ell_im_i}(\hk_i)\right]\P_{L_1L_2(L')L_3L_4}(\hk_1,\hk_2,\hk_3,\hk_4)\P_{l_1l_2(l')l_3l_4}(\hk_1,\hk_2,\hk_3,\hk_4)\\\nonumber
	T^{\ell_1\ell_2\ell_3\ell_4}_{m_1m_2m_3m_4} &=&
	(4\pi)^{8}\frac{8}{3}\sqrt\frac{2}{5} A^{(\rm Ghost -II)}\sum_{L_1\cdots L_4L'}i^{\ell_{1234}-L_{1234}}\tjo{L_1}{L_2}{L'}\tjo{L'}{L_3}{L_4}\mathcal{C}_{L_1L_2L_3L_4L'}\\\nonumber
	&&\,\times\, \int_{0}^{\infty}\lambda^{13}d\lambda\,\int x^2dx\,h^{\ell_1L_1}_{0,1/2}(x,\lambda)h^{\ell_2L_2}_{0,5/2}(x,\lambda)h^{\ell_3L_3}_{0,3/2}(x,\lambda)h^{\ell_4L_4}_{1,1/2}(x,\lambda)\\\nonumber
	&&\,\times\,\\\nonumber
	&&\,\times\,\prod_{i=1}^4\left[\int d\hk_iY^*_{\ell_im_i}(\hk_i)\mathcal{T}_{\ell_i}(k_i)\right]\P_{L_1L_2(L')L_3L_4}(\hk_1,\hk_2,\hk_3,\hk_4)\P_{22(1)10}(\hk_1,\hk_2,\hk_3,\hk_4),
\eeq
making use of the definition
\beq
    h^{\ell L}_{\nu\alpha}(x,\lambda) = \int\frac{k^2dk}{2\pi^2}k^\alpha \mathcal{T}_{\ell}(k)j_L(kx)H_{3/4-\nu}^{(1)}(2ik^2\lambda^2).
\eeq
To perform the $\hk$ integrals in the final lines of \eqref{eq: ghost-Tl}, we use the following relation:
\beq\label{eq: Basis5Prod}
    &&\prod_{i=1}^4\left[\int d\hk_i\,Y_{\ell_im_i}^*(\hk_i)\right]\P_{L_1L_2(L')L_3L_4}(\hk_1,\hk_2,\hk_3,\hk_4)\P_{\lambda_1\lambda_2(\lambda')\lambda_3\lambda_4}(\hk_1,\hk_2,\hk_3,\hk_4)\\\nonumber
    &=&\frac{1}{(4\pi)^2}\sum_{LM}(2L+1)(-1)^{L-M}\tj{\ell_1}{\ell_2}{L}{m_1}{m_2}{-M}\tj{L}{\ell_3}{\ell_4}{M}{m_3}{m_4}\prod_{i=1}^4\left[\tjo{L_i}{\lambda_i}{\ell_i}\right]\\\nonumber
    &&\,\times\,\mathcal{C}_{\ell_1\ell_2\ell_3\ell_4}\mathcal{C}_{\lambda_1\lambda_2\lambda_3\lambda_4\lambda'}\mathcal{C}_{L_1L_2L_3L_4L'}\begin{Bmatrix}L_1 & L_2 & L'\\ \lambda_1&\lambda_2& \lambda'\\ \ell_1&\ell_2&L\end{Bmatrix}\begin{Bmatrix}L' & L_3 & L_4\\ \lambda'&\lambda_3& \lambda_4\\ L&\ell_3&\ell_4\end{Bmatrix},
\eeq
which follows from \citep[\S6.5]{Cahn:2020axu} and the orthogonality of spherical harmonics. The quantities in curly braces are Wigner $9j$ symbols. Crucially, this involves Wigner $3j$ symbols which are of the same form as those appearing in the reduced trispectrum definition, thus we can read-off the relevant trispectrum components.\footnote{Formally, one can show this by contracting both \eqref{eq: Basis5Prod} and \eqref{eq: reduced-Tl-def} with $m$-weighted $3j$ symbols. In the latter case, the RHS will be a set of reduced trispectra $t^{\ell_1'\ell_2'}_{\ell_3'\ell_4'}(L')$ (where the primed $\ell$s are some permutation of $\{\ell_1,\ell_2,\ell_3,\ell_4)$), multiplied by a matrix involving $6j$ symbols. The same linear operator appears for the theoretical trispectrum, and thus can be trivially inverted to find the above relation.}

In full, we find the following reduced CMB trispectra sourced by ghost inflation at leading and subleading order:
\beq\label{eq: ghost-tl}
    &&\tj{\ell_1}{\ell_2}{L}{-1}{-1}{2}\tj{\ell_3}{\ell_4}{L}{-1}{-1}{2}t^{\ell_1\ell_2}_{\ell_3\ell_4}(L) = 2(4\pi)^7\ A^{(\rm Ghost -I)}(-1)^L\sum_{l_1l_2l_3l_4l'}c_{l_1l_2(l')l_3l_4}\mathcal{C}_{l_1l_2l_3l_4l'}\sum_{L'}(2L'+1)\\\nonumber
    &&\qquad\,\times\,\mathrm{Im}\int x^2 dx\int d\lambda\, \lambda^{11}\\\nonumber
    &&\,\times\,\left[\sum_{L_1L_2}i^{\ell_{12}-L_{12}}(2L_1+1)(2L_2+1)\tjo{L_1}{L_2}{L'}\tjo{L_1}{l_1}{\ell_1}\tjo{L_2}{l_2}{\ell_2}\begin{Bmatrix}L_1 & L_2 & L'\\ l_1&l_2& l'\\ \ell_1&\ell_2&L\end{Bmatrix}h^{\ell_1L_1}_{0,1/2}(x,\lambda)h^{\ell_2L_2}_{0,3/2}(x,\lambda)\right]\\\nonumber
    &&\,\times\,\left[\sum_{L_3L_4}i^{\ell_{34}-L_{34}}(2L_3+1)(2L_4+1)\tjo{L_3}{L_4}{L'}\tjo{L_3}{l_3}{\ell_3}\tjo{L_4}{l_4}{\ell_4}\begin{Bmatrix}L' & L_3 & L_4\\ l'&l_3& l_4\\ L&\ell_3&\ell_4\end{Bmatrix}h^{\ell_3L_3}_{0,1/2}(x,\lambda)h^{\ell_4L_4}_{0,1/2}(x,\lambda)\right]\\\nonumber
    &&\tj{\ell_1}{\ell_2}{L}{-1}{-1}{2}\tj{\ell_3}{\ell_4}{L}{-1}{-1}{2}t^{\ell_1\ell_2}_{\ell_3\ell_4}(L) =
	-(4\pi)^{7}40\sqrt\frac{2}{15} A^{(\rm Ghost -II)}(-1)^L\sum_{L'}(-1)^{L'}(2L'+1)\int x^2dx\int d\lambda\,\lambda^{13}\\\nonumber
	&&\qquad\,\times\,\left[\sum_{L_1L_2}i^{\ell_{12}-L_{12}}(2L_1+1)(2L_2+1)\tjo{L_1}{L_2}{L'}\tjo{L_1}{2}{\ell_1}\tjo{L_2}{2}{\ell_2}\begin{Bmatrix}L_1 & L_2 & L'\\ 2&2& 1\\ \ell_1&\ell_2&L\end{Bmatrix}h^{\ell_1L_1}_{0,1/2}(x,\lambda)h^{\ell_2L_2}_{0,5/2}(x,\lambda)\right]\\\nonumber
	&&\qquad\,\times\,\left[\sum_{L_3}i^{-\ell_3-L_3}(2L_3+1)\tjo{L_3}{\ell_4}{L'}\tjo{L_3}{1}{\ell_3}\begin{Bmatrix} L' & L_3 & \ell_4 \\ \ell_3 & L & 1\end{Bmatrix}h^{\ell_3L_3}_{0,3/2}(x,\lambda)h^{\ell_4\ell_4}_{1,1/2}(x,\lambda)\right],
\eeq
where we have noted that the $(1,2)$ and $(3,4)$ components separate, allowing for efficient $\ell$-space binning, and computation via summation and two coupled integrals (given that the Hankel functions are smooth in $k\lambda$). In practice, the above form is computed by first transforming to the variables $q=k\lambda$ and $y=x/\lambda$ (as in \citep{Cabass:2022oap}), then evaluating the integral using numerical quadrature in $q$, $\log y$, and $\log \lambda$.

\subsection{Cosmological Collider}
Next, we consider inflationary signals from the `Cosmological Collider' mechanism; in particular, the exchange of a spin-1 particle, $\sigma_\mu$, with mass $m_\sigma$ and sound-speed $c_s$. As shown in \citep{Cabass:2022rhr,Cabass:2022oap}, the pre-symmetrized particle-exchange trispectrum can be written
\beq\label{eq: collider-Tk}
    \tilde{T}_{\lambda_1\lambda_3}(\vk_1,\vk_2,\vk_3,\vk_4) = -\left[ic_s^4\frac{\lambda_1\lambda_3}{2H}(\Delta_\zeta^2)^4\sin\pi(\nu+1/2)\right](\hk_1\cdot\hk_2)(\hk_3\cdot\hk_4)(\hk_2\cdot\hk_3\times\hk_4)t^A(k_1,k_2,s)t^B(k_3,k_4,s),
\eeq
where $\vs=\vk_1+\vk_2$ is a Mandelstam variable, $\lambda_1$ and $\lambda_3$ are the interaction strengths,  and $\nu = \sqrt{9/4-m_\sigma^2/H^2}$. We additionally define
\beq
    t^A(k_1,k_2,s) &=& k_1^{-2}k_2^{-1}(k_1-k_2)\left[k_{12}J_3(c_sk_{12},s)+c_sk_1k_2J_4(c_sk_{12},s)\right]\\\nonumber
    t^B(k_3,k_4,s) &=& k_3^{-1}k_4^{-1}(k_3-k_4)\left[k_{34}J_4(c_sk_{34},s)+c_sk_3k_4J_5(c_sk_{34},s)\right]
\eeq
with $k_{12}\equiv k_1+k_2$ and $k_{34}\equiv k_3+k_4$, Since $\hk_1\cdot(\hk_3\times\hk_4)=-\hk_2\cdot(\hk_3\times\hk_4)$, \eqref{eq: collider-Tk} is symmetric under $\vk_1\leftrightarrow\vk_2$. This involves the $J_n$ functions, defined via
\beq
    J_n(a,b) = \left(\frac{1}{2b} \right)^{n+1/2} \frac{\Gamma \left(\alpha \right)\Gamma \left(\beta \right)}{\Gamma(1+n)} \,\, {}_2 F_1 \Big(\alpha, \beta; 1+n; \frac{1}{2}-\frac{a}{2b}\Big),
\eeq
where $\alpha = 1/2+n-\nu$, $\beta = 1/2+n+\nu$. In this case, we will constrain the amplitude
\beq\label{eq: A-collider}
    A^{(\rm Collider)}(c_s,\nu) = \frac{\lambda_1\lambda_3}{H} c_s^4(\Delta_\zeta^2)^4\sin\pi(\nu+1/2),
\eeq
as in \citep{Cabass:2022oap}.

Since the above trispectrum is of the exchange form it is useful to rewrite the Dirac delta as an integral over the exchange momentum $\vs$:
\beq\label{eq: Dirac-s-expansion}
    &&\delD{\vk_{1234}} = \int_{\vs}\delD{\vk_{12}-\vs}\delD{\vk_{34}+\vs}\\\nonumber
    &&= (4\pi)^4\sum_{L_1L_2L_3L_4L'}(-i)^{L_{1234}}\mathcal{C}_{L_1L_2L_3L_4L'}\tjo{L_1}{L_2}{L'}\tjo{L_3}{L_4}{L'}\P_{L_1L_2(L')L_3L_4}(\hk_1,\hk_2,\hk_3,\hk_4)\\\nonumber
    &&\qquad\times\,\int\frac{s^2ds}{2\pi^2}\left[\int_0^{\infty}x^2dx\,j_{L_1}(k_1x)j_{L_2}(k_2x)j_{L'}(sx)\right]\left[\int_0^{\infty}y^2dy\,j_{L_3}(k_3y)j_{L_4}(k_4y)j_{L'}(sy)\right],
\eeq
where the second line is derived in \citep[Eq.\,A9]{Cabass:2022oap}. We note that the infinite integrals over $x$ and $y$ can be written as a finite sum over Wigner $6j$ symbols, following \citep{Mehrem_1991}.

Secondly, the angular parts of \eqref{eq: collider-Tk} can be expanded as \citep[Eq.\,A21]{Cabass:2022oap}:
\beq
    (\hk_1\cdot\hk_2)(\hk_3\cdot\hk_4)(\hk_2\cdot\hk_3\times\hk_4) = i\frac{(4\pi)^2}{9\sqrt 5}\left[2\,\P_{12(1)22}(\hk_1,\hk_2,\hk_3,\hk_4)-\sqrt{2}\,\P_{10(1)22}(\hk_1,\hk_2,\hk_3,\hk_4)\right].
\eeq
Finally, the two basis functions can be integrated against the $Y_{\ell_im_i}^*(\hk_i)$ harmonics using the relation \eqref{eq: Basis5Prod}, and the reduced trispectrum coefficients read off as before. Combining results, we find the trispectrum form
\beq\label{eq: reduced-T-collider}
    &&\tj{\ell_1}{\ell_2}{L}{-1}{-1}{2}\tj{\ell_3}{\ell_4}{L}{-1}{-1}{2}t^{\ell_1\ell_2}_{\ell_3\ell_4}(L) = \frac{\sqrt{5}}{6}(4\pi)^{5}A^{(\rm Collider)}(c_s,\nu)(-1)^{L}\sum_{L'}\int\frac{s^2ds}{2\pi^2}\\\nonumber
    &&\qquad\qquad\qquad\times\,\left[\sum_{L_1L_2}i^{\ell_{12}-L_{12}}\mathcal{M}^{\ell_1\ell_2}_{L_1L_2}(L,L')Q^{\ell_1\ell_2,A}_{L_1L_2L'}(s)\right]\left[\sum_{L_3L_4}i^{\ell_{34}-L_{34}}\mathcal{N}^{\ell_3\ell_4}_{L_3L_4}(L,L')Q^{\ell_3\ell_4,B}_{L_3L_4L'}(s)\right],
\eeq
which we have expressed in terms of two pieces, connected only by $L'$ and $s$ (\textit{i.e.}\ the exchange variables). This uses the definitions
\beq
    \mathcal{M}^{\ell_1\ell_2}_{L_1L_2}(L,L') &=&\mathcal{C}^2_{L_1L_2}\tjo{L_1}{L_2}{L'}\tjo{L_1}{1}{\ell_1}\\\nonumber
    &&\,\times\,\left[2\sqrt{5}\tjo{L_2}{2}{\ell_2}\begin{Bmatrix}L_1 & L_2 & L'\\ 1&2&1\\ \ell_1&\ell_2&L\end{Bmatrix}+\sqrt{\frac{2}{3}}\frac{\delta^{\rm K}_{\ell_2L_2}}{2\ell_2+1}(-1)^{\ell_1+L'}\begin{Bmatrix}L&\ell_1&\ell_2\\L_1&L'&1\end{Bmatrix}\right]\\\nonumber
    \mathcal{N}^{\ell_3\ell_4}_{L_3L_4}(L,L') &=&
    \mathcal{C}^2_{L_3L_4L'}\tjo{L_3}{L_4}{L'}\tjo{L_3}{2}{\ell_3}\tjo{L_4}{2}{\ell_4}\begin{Bmatrix}1&2&2\\L' & L_3 & L_4\\L&\ell_3&\ell_4\end{Bmatrix},
\eeq
and 
\beq
     Q^{\ell_1\ell_2,X}_{L_1L_2L'}(s) = \int\frac{k_1^2dk_1}{2\pi^2}\frac{k_2^2dk_2}{2\pi^2}\mathcal{T}_{\ell_1}(k_1)\mathcal{T}_{\ell_2}(k_2)\int_0^{\infty}x^2dx\,j_{L_1}(k_1x)j_{L_2}(k_2x)j_{L'}(sx)\,t^X(k_1,k_2,s).
\eeq
Practically, we recast the $Q$ integrals in the rescaled variables $K_{\pm}=(k_1\pm k_2)/(2s)$ and perform the $x$ integral analytically using \citep{Mehrem_1991}, noting that triangle conditions on $\{L_1,L_2,L'\}$ restrict the range of $K_{\pm}$.\footnote{At large $K_i$, this approach becomes numerically unstable, thus we switch to explicit numerical integration.} The latter yields a function $f(K_+,K_-)$, which can be efficiently combined with the rescaled trispectrum components $s^{-3/2}t^X(K_++K_-,K_+-K_-,1)$, and transfer functions $\mathcal{T}_{\ell_i}(K_+s\pm K_-s)$.

\subsection{Chern-Simons Gauge Fields}
Finally, we consider the inflationary model of \citep{Shiraishi:2016mok}, involving the exchange $U(1)$ gauge fields in a Chern-Simons background \resub{(see also \citep{Niu:2022fki})}. From \citep{Philcox:2022hkh}, we have the pre-symmetrized trispectrum:
\beq
    \tilde{T}_{\rm CS}(\vk_1,\vk_2,\vk_3,\vk_4) = 3iA^{(\rm Gauge)}(\Delta_\zeta^2)^3k_1^{-3}k_3^{-3}s^{-3}\left[1-\hk_1\cdot\hk_3+\hk_1\cdot\hs-\hk_3\cdot\hs\right]\left[\hs\cdot(\hk_1\times\hk_3)\right],
\eeq
for characteristic amplitude $A^{(\rm Gauge)}$, related to microphyscial parameters as:
\beq\label{eq: A-gauge}
    A^{(\rm Gauge)} \approx \frac{5\times 10^3}{|\epsilon|}\frac{e^{12\pi|\gamma|}}{|\gamma|^9}\left(\frac{N}{60}\right)^4\frac{\rho_E}{\rho_\phi},
\eeq
and equal to the $A_{\rm CS}$ parameter of \citep{Philcox:2022hkh}. Here $\gamma$ is the Chern-Simons coupling, $N$ is the number of $e$-folds of inflation, $\rho_E/\rho_\phi$ is the fractional energy density in the Gauge field and $\epsilon$ is the slow-roll parameter. Since this involves particle exchange, we expand the corresponding Dirac delta in a manner akin to \eqref{eq: Dirac-s-expansion}, but do not yet integrate over $\hs$ (cf.\,\citep[Eq.\,B.3]{Philcox:2022hkh})
\beq
    &&\delD{\vk_{1234}} = \int_{\vs}\delD{\vk_{12}-\vs}\delD{\vk_{34}+\vs}\\\nonumber
    &&= (4\pi)^5\int_{\vs}\sum_{L_1L_2L_3L_4L'L''}i^{L_{1234}-L'+L''}\mathcal{C}_{L_1L_2L_3L_4L'L''}\tjo{L_1}{L_2}{L'}\tjo{L_3}{L_4}{L''}\P_{L_1L_2L'}(\hk_1,\hk_2,\hs)\P_{L_3L_4L''}(\hk_3,\hk_4,\hs)\\\nonumber
    &&\qquad\times\,\int_0^{\infty}x^2dx\,j_{L_1}(k_1x)j_{L_2}(k_2x)j_{L'}(sx)\int_0^{\infty}y^2dy\,j_{L_3}(k_3y)j_{L_4}(k_4y)j_{L''}(sy),
\eeq
For the angular components, we note that \citep[Eq.\,B1]{Philcox:2022hkh}\footnote{We correct a pair of sign errors in the former work.}
\beq
    3i\left[1-\hk_1\cdot\hk_3+\hk_1\cdot\hs-\hk_3\cdot\hs\right]\left[\hs\cdot(\hk_1\times\hk_3)\right] = -\sqrt{2}(4\pi)^{3/2}\left[\P_{111}+\frac{1}{\sqrt{5}}\P_{221}-\frac{1}{\sqrt{5}}\P_{212}+\frac{1}{\sqrt{5}}\P_{122}\right](\hk_1,\hk_3,\hs)
\eeq

In combination, the CMB trispectrum takes the form
\beq
    T^{\ell_1\ell_2\ell_3\ell_4}_{m_1m_2m_3m_4} &=& -\sqrt{2}i^{\ell_{1234}}A^{(\rm Gauge)}(\Delta_\zeta^2)^3(4\pi)^{11/2}
    \sum_{L_1L_2L_3L_4L'L''}i^{L_{1234}-L'+L''}\mathcal{C}_{L_1L_2L_3L_4L'L''}\tjo{L_1}{L_2}{L'}\tjo{L_3}{L_4}{L''}\\\nonumber
    &&\,\times\,\\\nonumber
    &&\qquad\times\,\int x^2dx\,\int y^2dy\,h^{\ell_1L_1}_{-3}(x)h^{\ell_2L_2}_{0}(x)h^{\ell_3L_3}_{-3}(y)h^{\ell_4L_4}_{0}(y)f_{L'L''}(x,y)\\\nonumber
    &&\int d\hs\,\prod_{i=1}^4\left[\int d\hk_iY^*_{\ell_im_i}(\hk_i)\right]\P_{L_1L_2L'}(\hk_1,\hk_2,\hs)\P_{L_3L_4L''}(\hk_3,\hk_4,\hs)\sum_{\lambda_1\lambda_3\lambda}A_{\lambda_1\lambda_3\lambda}\P_{\lambda_1\lambda_3\lambda}(\hk_1,\hk_3,\hs)+\text{23 perms.},
\eeq
now defining 
\beq
    h^{\ell L}_{\alpha}(x)&=&\int\frac{k^2dk}{2\pi^2}k^\alpha\mathcal{T}_{\ell}(k)j_L(kx), \qquad f_{LL'}(x,y) =\int\frac{s^2ds}{2\pi^2}\frac{j_L(sx)j_{L'}(sy)}{s^3}
\eeq
with $A_{111}=1$, $A_{221}=-A_{212}=A_{122}=1/\sqrt{5}$ and zero else.

To proceed, we must perform the angular integral. This is a lengthy calculation, which we do not quote here to avoid unnecessary tedium, but yields the following form:
\beq
    &&\int d\hs\,\prod_{i=1}^4\left[\int d\hk_iY^*_{\ell_im_i}(\hk_i)\right]\P_{L_1L_2L'}(\hk_1,\hk_2,\hs)\P_{L_3L_4L''}(\hk_3,\hk_4,\hs)\P_{\lambda_1\lambda_3\lambda}(\hk_1,\hk_3,\hs)\\\nonumber
    &&=\delta^{\rm K}_{\ell_2L_2}\delta^{\rm K}_{\ell_4L_4}(4\pi)^{-3/2}\mathcal{C}_{\lambda_1\lambda_3\lambda}\mathcal{C}_{L_1L_3L'L''}\mathcal{C}_{\ell_1\ell_3LL}\sum_{LM}\tjo{L_1}{\lambda_1}{\ell_1}\tjo{L'}{\lambda}{L''}\tjo{\lambda_3}{L_3}{\ell_3}\\\nonumber
    &&\,\times\,(-1)^{\ell_1+L''+L_3}\begin{Bmatrix} L&\ell_1&\ell_2\\L_1&L'&\lambda_1\end{Bmatrix}\begin{Bmatrix}\lambda&\lambda_1&\lambda_3\\L&L''&L'\end{Bmatrix}\begin{Bmatrix}\ell_3&\ell_4&L\\L''&\lambda_3&L_3\end{Bmatrix}(-1)^{M}\tj{\ell_1}{\ell_2}{L}{m_1}{m_2}{-M}\tj{L}{\ell_3}{\ell_4}{M}{m_3}{m_4}
\eeq
This yields the reduced trispectrum
\beq
    t^{\ell_1\ell_2}_{\ell_3\ell_4}(L)\tj{\ell_1}{\ell_2}{L}{-1}{-1}{2}\tj{\ell_3}{\ell_4}{L}{-1}{-1}{2} &=& -\sqrt{2}A^{(\rm Gauge)}(\Delta_\zeta^2)^3(4\pi)^5    \sum_{L_1L_3L'L''}i^{-\ell_1+\ell_3-L_1+L_3+L'+L''}\mathcal{C}_{L_1L_3L'L''}^2\sum_{\lambda_1\lambda_3\lambda}A_{\lambda_1\lambda_3\lambda}\mathcal{C}_{\lambda_1\lambda_3\lambda}\nonumber\\
    &&\,\times\,\int x^2dx\,\int y^2dy\,h^{\ell_1L_1}_{-3}(x)h^{\ell_2\ell_2}_{0}(x)h^{\ell_3L_3}_{-3}(y)h^{\ell_4\ell_4}_{0}(y)f_{L'L''}(x,y)\\\nonumber
    &&\,\times\,\tjo{L_1}{\ell_2}{L'}\tjo{L_3}{\ell_4}{L''}\tjo{L_1}{\lambda_1}{\ell_1}\tjo{L'}{\lambda}{L''}\tjo{\lambda_3}{L_3}{\ell_3}\\\nonumber
    &&\,\times\,\begin{Bmatrix} L&\ell_1&\ell_2\\L_1&L'&\lambda_1\end{Bmatrix}\begin{Bmatrix}\lambda&\lambda_1&\lambda_3\\L&L''&L'\end{Bmatrix}\begin{Bmatrix}\ell_3&\ell_4&L\\L''&\lambda_3&L_3\end{Bmatrix},
\eeq
noting that, due to the $3j$ symbols and the restricted $\lambda_i$ range, $L_i\in[\ell_i-2,\ell_i+2]$, $|L'-L''|\leq 2$ and $L'\leq 2\ell_2$.

This is more conveniently expressed in factorized form
\beq\label{eq: gauge-tl}
    t^{\ell_1\ell_2}_{\ell_3\ell_4}(L)\tj{\ell_1}{\ell_2}{L}{-1}{-1}{2}\tj{\ell_3}{\ell_4}{L}{-1}{-1}{2} &=& -\sqrt{2}A^{(\rm gauge)}(\Delta_\zeta^2)^3(4\pi)^5\sum_{\lambda_1\lambda_3\lambda}A_{\lambda_1\lambda_3\lambda}\mathcal{C}_{\lambda_1\lambda_3\lambda}\\\nonumber
    &&\,\times\,\sum_{L'L''}(2L'+1)(2L''+1)\tjo{L'}{\lambda}{L''}\begin{Bmatrix}\lambda&\lambda_1&\lambda_3\\L&L''&L'\end{Bmatrix}\int\frac{s^2ds}{2\pi^2}\frac{1}{s^3}\\\nonumber
    &&\,\times\,\left[\sum_{L_1}i^{-\ell_1-L_1+L'}(2L_1+1)\tjo{L_1}{\ell_2}{L'}\tjo{L_1}{\lambda_1}{\ell_1}\begin{Bmatrix} L&\ell_1&\ell_2\\L_1&L'&\lambda_1\end{Bmatrix}\right.\\\nonumber
    &&\,\qquad\qquad\,\left.\times\,\int x^2dx\,h^{\ell_1L_1}_{-3}(x)h^{\ell_2\ell_2}_{0}(x)j_{L'}(sx)\right]\\\nonumber
    &&\,\times\,
  \left[\sum_{L_3}i^{\ell_3+L_3+L''}(2L_3+1)\tjo{L_3}{\ell_4}{L''}\tjo{\lambda_3}{L_3}{\ell_3}\begin{Bmatrix}\ell_3&\ell_4&L\\L''&\lambda_3&L_3\end{Bmatrix}\right.\\\nonumber
    &&\,\qquad\qquad\,\left.\times\,\int y^2dy\,h^{\ell_3L_3}_{-3}(y)h^{\ell_4\ell_4}_{0}(y)j_{L''}(sy)\right],
\eeq
where the two pieces are coupled by the exchange momentum $s$. This reduces the number of nested summations (since there are no cross-terms in $(L_1,L_3)$) and, further, one can directly apply binning to the two brackets (after accounting for the $3j$ prefactors). In practice, the integrals are evaluated using numerical quadrature, with linearly spaced arrays in $\{k_i/\ell_i,x,s\}$ (noting that the $k$ integrals can all be precomputed).

In the Sachs-Wolfe limit, $\mathcal{T}_\ell(k) = -(1/5)j_\ell(kD_\ast)$ (for distance $D_\ast$ to last scattering), which gives the simplification
\beq
    h^{\ell\ell}_0(x) &=& -\frac{1}{20\pi D_\ast^2}\delta_{\rm D}(x-D_\ast)\\\nonumber
    -5h^{LL'}_{-3}(D_\ast) = f_{LL'}(D_\ast,D_\ast) &=& \frac{\Gamma\left(\tfrac{L+L'}{2}\right)}{16\pi\,\Gamma\left(\tfrac{L-L'+3}{2}\right)\Gamma\left(\tfrac{L'-L+3}{2}\right)\Gamma\left(\tfrac{L+L'+4}{2}\right)}
\eeq
(using \citep[Eq.\,30]{Shiraishi:2016mok}). This allows the large-scale trispectrum template to be computed without any integrals. The same does not hold for the Collider or Ghost templates, since these have more complex dependence on $k$.

\end{document}